\documentclass[journal, 10pt]{IEEEtran}
\usepackage{algorithm,multirow}
\usepackage[noend]{algorithmic}
\usepackage{graphicx}
\usepackage{comment}
\usepackage{cite}
\usepackage{multirow}
\usepackage{amssymb,amsthm,mathrsfs}
\usepackage{amsfonts,epsf,graphicx,times,amsmath,multirow,url,cite}
\usepackage{float}
\usepackage{flexisym}
\usepackage[capitalise]{cleveref}
\usepackage{float}
\usepackage{verbatim}

\usepackage{amsthm}
\usepackage{subcaption}
\usepackage{color}

\usepackage{footnote}
\usepackage{footmisc}
\makesavenoteenv{tabular}
\makesavenoteenv{table}
\renewcommand{\thefootnote}{\alph{footnote}}
\newcommand{\astfootnote}[1]{
\let\oldthefootnote=\thefootnote
\setcounter{footnote}{0}
\renewcommand{\thefootnote}{\fnsymbol{footnote}}
\footnote{#1}
\let\thefootnote=\oldthefootnote
}

\usepackage{wrapfig}

\graphicspath{{figs/}}
	\DeclareGraphicsExtensions{.eps}

\usepackage{array}
\newcolumntype{L}[1]{>{\raggedright\let\newline\\\arraybackslash\hspace{0pt}}m{#1}}
\newcolumntype{C}[1]{>{\centering\let\newline\\\arraybackslash\hspace{0pt}}m{#1}}
\newcolumntype{R}[1]{>{\raggedleft\let\newline\\\arraybackslash\hspace{0pt}}m{#1}}

\makeatletter
\def\ps@headings{%
\def\@oddhead{\mbox{}\scriptsize\rightmark \hfil \thepage}%
\def\@evenhead{\scriptsize\thepage \hfil \leftmark\mbox{}}%
\def\@oddfoot{}%
\def\@evenfoot{}}
\makeatother
\makeatletter

\usepackage{comment}
\let\wfs@comment@comment\comment
\let\comment\@undefined

\usepackage[final]{changes}
\let\wfs@changes@comment\comment
\let\comment\@undefined

\newcommand\comment{%
    \ifthenelse{\equal{\@currenvir}{comment}}
    {\wfs@comment@comment}
    {\wfs@changes@comment}%
}

\makeatother

\usepackage [english]{babel}
\usepackage [autostyle, english = american]{csquotes}
\MakeOuterQuote{"}
\definechangesauthor[name={Jose Terrero}, color=red]{JT}
\definechangesauthor[name={Yuqi Fu}, color=green]{yuqi}

\usepackage{caption}
\usepackage{subcaption}

\newcommand{\sol}{\texttt{SpeCon}}

\begin{document}

\date{}


\title{\huge Speculative Container Scheduling for Deep Learning Applications in a Kubernetes Cluster}
\author{

Ying Mao,~\IEEEmembership{Member,~IEEE,}  Yuqi Fu, Wenjia Zheng,~\IEEEmembership{Student Member,~IEEE,} 
\\Long Cheng, Qingzhi Liu, and Dingwen Tao~\IEEEmembership{Member,~IEEE}

\IEEEcompsocitemizethanks{

\IEEEcompsocthanksitem Y. Mao, Y. Fu, and W. Zheng are with the Department of Computer and Information Science at Fordham University in the New York City. E-mail: \{ymao41, yfu81, wzheng33\}@fordham.edu
\IEEEcompsocthanksitem L. Cheng is with the School of Computing, Dublin City University, Ireland. E-mail: long.cheng@dcu.ie
\IEEEcompsocthanksitem Q. Liu is with Information Technology Group, Wageningen Univer- sity, The Netherlands. E-mail: qingzhi.liu@wur.nl
\IEEEcompsocthanksitem D. Tao is with School of Electrical Engineering and Computer Science at Washington State University, Pullman, Washington. E-mail: dingwen.tao@ieee.org

}
}

\maketitle

\begin{abstract}

In the past decade, we have witnessed a dramatically increasing volume of data collected from varied sources.
The explosion of data has transformed the world as more information is available for collection and analysis than ever before.
To maximize the utilization, various machine and deep learning models have been developed, e.g. CNN~\cite{cnn} and RNN~\cite{rnn}, to study data and extract valuable information from different perspectives. 
While data-driven applications improve countless products,
training models for hyperparameter tuning is still a time-consuming and resource-intensive process. 
Cloud computing provides infrastructure support for the training of deep learning applications. The cloud service providers,
such as Amazon Web Services~\cite{aws},
create an isolated virtual environment (virtual machines and containers) for clients, who share physical resources, e.g., CPU and memory. On the cloud, resource management schemes are implemented to enable better sharing among users and boost the system-wide performance. However, general scheduling approaches, such as spread priority and balanced resource schedulers, do not work well with deep learning workloads. In this project, we propose \sol~, a novel container scheduler that is optimized 
for short-lived deep learning applications.
Based on virtualized containers, such as Kubernetes~\cite{k8s} and Docker~\cite{docker}, \sol~ analyzes the common characteristics of training processes.
We design a suite of algorithms to monitor the progress of the training and speculatively migrate the slow-growing models to release resources for fast-growing ones. Specifically, the extensive experiments demonstrate that \sol~ improves the completion time of an individual job by up to 41.5\%, 14.8\% system-wide and 24.7\% in terms of makespan.

\end{abstract}



\section{Introduction}
With the dramatic growth of data volume from different sources, such as Edge, Cloud and Internet of Things~\cite{acharya2019edge, acharya2019workload, harvey2017edos, mao2017draps, mao2014pasa},
tons of businesses have transformed themselves to embrace data driven models. 
A number of deep learning and data mining backed applications are reforming our daily life. For instance, Uber~\cite{uber} and Lyft~\cite{lyft} utilize the data to predicate the traffic and demand, which benefits clients with sharing economies. Another example is Apple FaceID, which uses computer vision to provide hands-free unlocking ability.
Leveraging the power of data, practitioners and researchers have proposed various learning architectures and models. Generative Adversarial Networks~\cite{gans} , for example, are taught to create words similar to given data sets in any domain: 
images, music, speech,and prose. A trained Generative Adversarial Network is a robot artist in a sense, who can compose impressive  work.

To obtain a well-trained model, an iterative optimization algorithm is usually involved in hyperparameter tuning in order to find the best parameter combination, which can optimize a pre-defined evaluation function. This function mathematically determines the performance of a model, such as classification accuracy and inception score~\cite{evaluation}. 
In each iteration, the model learns a batch of samples (training data). Based on the training results, the parameters in the model will be updated according to the evaluation function.

Despite the excellent performance achieved by learning models, the training process is both time-consuming and resource intensive.
Depending on the hardware types, training the database of ImageNet~\cite{imagenet} may take hours or even days~\cite{you2018imagenet}. Consequently, the process of training models is usually done on the cloud, where big companies, like Amazon Web Services~\cite{aws}, Microsoft Azure~\cite{azure} and Google Cloud Platform~\cite{gcp}, provide infrastructures support.
The service providers create an isolated computing environment with traditional virtual machines and/or more recently, virtual containers, such as Kubernetes~\cite{k8s} and Docker~\cite{docker}. In the cloud, resources, such as CPU and memory, are shared by many users. Therefore, cloud providers implement scheduling algorithms from different perspectives to ensure a better sharing of resources. For example, Kubernetes provides multiple options~\cite{k8s-scheduler} that can be adopted to the system scheduler, such as Balanced Resource Allocation, which favors nodes with balanced resource usage, and Selector Spread Priority, which spreads the workload across hosts. 

These general scheduling approaches, however, do not work well for training processes of deep learning applications. First of all, they fail to consider characteristics of evaluation functions, where the loss reduces dramatically at the beginning of the process, then converges and becomes stable. Although the resource usage may stay at the same level, the gain of the training process varies over the time. Secondly, the schedulers, in current system, migrate or terminate a container when it consumes much more resources than others and overwhelmes the physical machine. With deep learning workloads, a converged model may hurt other training jobs on the same machine since the scheduler fails to distinguish them.

With respects to the features of deep learning applications, we develop \sol, a Speculative Container scheduler, to accelerate the training jobs in a containerized cluster shared by multiple users. Based on the growth of evaluation functions, \sol~ categorizes a training job into different phases, which reflects the dynamics of the efficiency. According to the categories, \sol~ decides when a deep learning model starts its convergence. For a converged job, \sol~ first gathers the information of the system-wide resource usages as well as the existing jobs on each node and then, migrates this job to a most desirable host at the moment. The container reallocation in \sol~ releases resources to fast-growing jobs and boost the overall system performance. When all containers are converged, \sol~ keeps monitoring the cluster and rebalancing the workloads.
The main contributions of this paper are summarized as follows.

\begin{itemize}

\item We analyze the characteristics of training processes for the various deep learning models and 
conduct experiments to study the overhead of saving and resuming jobs in a Kubernetes cluster.

\item We propose \sol~ with a suite of algorithms that monitor the evaluation functions of deep learning models on the worker side and on the manager side. It collects system-wide information to calculate a weighted score for each worker to select a most desirable host for the migration. Furthermore, it rebalances the workload when all containers are converged.

\item We implement \sol~ on top of a leading container orchestration platform, Kubernetes, and evaluate it with popular containerized deep learning applications. With intensive experiments on the cloud, we demonstrate that \sol~ achieves significant improvement in the time of completion. It reduces an individual job by as large as 41.5\%, 14.8\% in average, 24.7\% for system makespan.

\end{itemize}


\section{Related Work}

With prevalence of data-driven businesses, learning from big data draws an increasing 
attention in both industry and academia. In this domain, various data mining algorithms
as well as deep learning frameworks have been proposed to maximize the utilization.
On the back-end side, the cloud computing powered all those platforms by providing on-demand
storage, computing and analytics services~\cite{saiyeda2017cloud, mao2020resource,chen2020woa, zheng2019target}.

The cloud providers create a virtualized and isolated computing environment for each of the clients, who
share resources on physical machines, such as CPU and memory. Traditionally, technologies of virtual machines 
are used for creating the environment~\cite{galante2012survey}. To start a virtual machine, the system is required to
boot its guest operating system, which consumes lots of resources and significantly increases the  provisioning delay~\cite{zhang2015guaranteeing}. In the literature, various approaches focused on improving the provision time for virtual machines
~\cite{xavier2016time, wu2016energy, reguri2016energy, mandal2016bandwidth, zhang2017cloudgc, sung2019virtual}.
While many efforts have been made in this field, reducing the provision cost remains a challenge task due to the 
requirement of guest operating systems.
After the provision, it is a common scenario that a live virtual machine need to be migrated to another node in the cluster to the accommodate the real-time network topology and workload~\cite{ahmad2015survey, noshy2018optimization, he2019performance}.
However, the heavy-weight design of storage system in virtual machines made the migration process an expensive operation. It usually cost a few tens of seconds for a regular virtual machine~\cite{zhang2018survey}.

To address the limitations, containers are designed for deploying and running distributed applications, where multiple isolated containers are sharing the host operating system and physical resources. 
A comparative study in~\cite{zhang2018comparative} demonstrated that containers, when compared to virtual machines, require far less provisioning and migration time. 
As the popularity increases,  container scheduling has become an emerging research topic. 
While Stratus~\cite{chung2018stratus} exploited cloud properties and runtime estimates to reduce the dollar cost of cluster jobs executed on public clouds, PIVOT~\cite{jiangpivot} investigated a cloud-agnostic system and the proposed scheduler enables data-intensive applications to scale across clouds instantly in a cost-efficient manner.
Focuing on heterogeneities of a cluster,  ECSched~\cite{hu2018ecsched} enabled high-quality and fast placement decisions for concurrent deployment requests on heterogeneous clusters. The ECSched mapped scheduling decisions to a graphic data structure and model it as minimum cost flow problem. Besides CPU-based clusters, authors in~\cite{thinakaran2018curious} explored a GPU rich datacenter and proposed Knots, a GPU-aware resource orchestration layer and integrate it with Kubernetes container orchestrator. 

Despite the efforts promoted in this field, limited number of them focused on scheduling containers for training processes of deep learning applications. While FlowCon~\cite{Zhengflowcon} tried to set limits to container's resource requests, the evaluation focused on the single machine setting. Furthermore, Gandiva~\cite{xiao2018gandiva} was developed as a new cluster scheduling framework that utilizes domain-specific knowledge to improve latency and efficiency of training models. However, it does not take into account the properties of evaluation functions of learning models. Additionally, ProCon~\cite{fuprogress} tried to 
solve container placement in Kubernetes, but it fails to take container migration into consideration and 
SLAQ~\cite{slaq} schedules concurrent
machine learning training jobs based on quality improvement for
resource usage, by allocating cluster resources iteratively. However, it fails to re-allocate the resources when workload updates.

We present \sol, in this project, which keeps tracking the convergence of the deep learning models based on their pre-defined evaluation functions. In \sol, scheduling decisions are made according to the individual gains of each model in a given period. The scheduler considers reallocation of a container when it becomes a slow-growing one.


\section{Background and Motivation}

In this section, we study the training process of containerized deep learning applications and motivate our work with an example.

\begin{figure}[ht]
\vspace{-0.1in}
\centering
\includegraphics[width=0.7\linewidth]{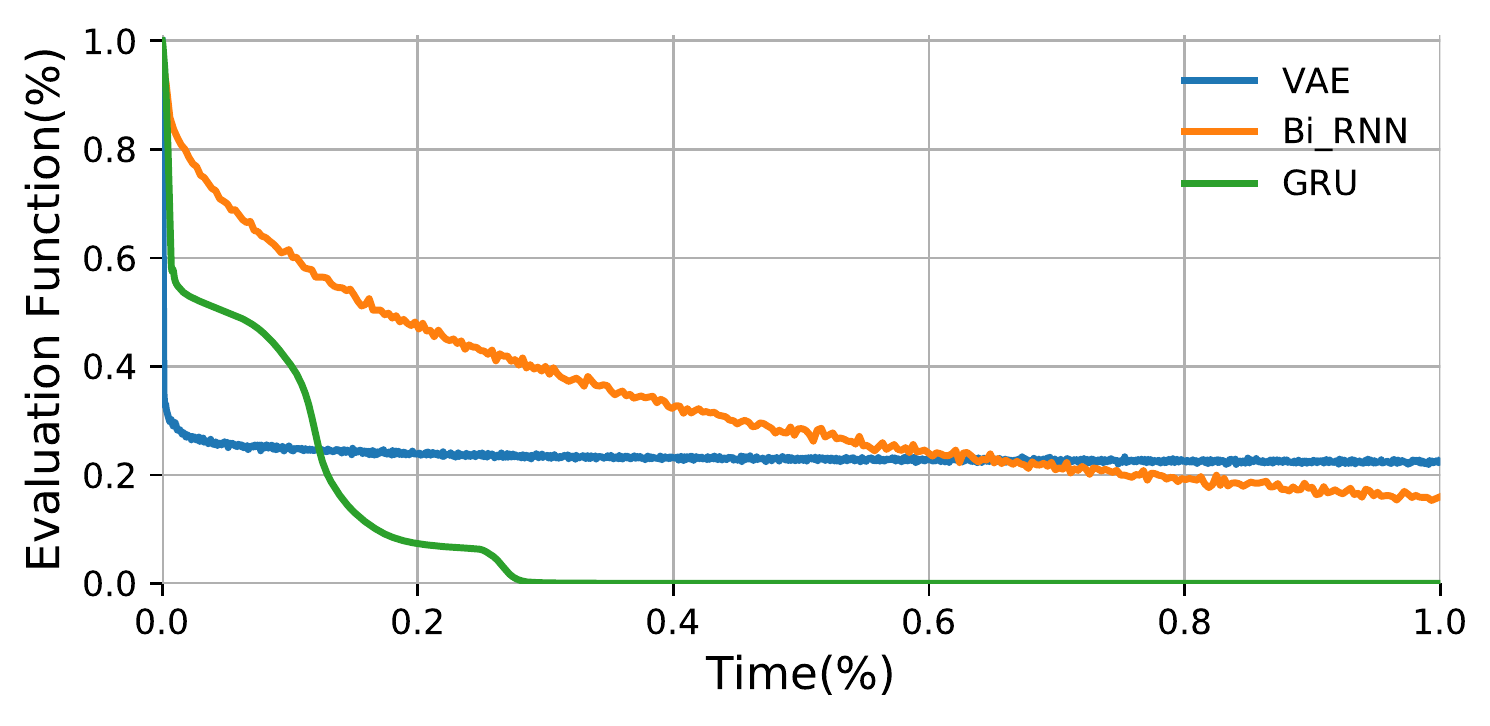}
\caption{Training progress of three deep learning models}
\label{fig:motivation}
\vspace{-0.1in}
\end{figure}

\subsection{Training a Deep Learning Model}

Given a deep learning algorithm, training the model is the parameters tuning process to approach the global optimum of
a predefined evaluation function. Initially, parameters are randomly selected for the model. This model is iteratively fed with 
mini-batches of training data, where each mini-batch contains features and labels. 
It is, then, evaluated with the evaluation function that informs users how far off does it from the target.
The returned value of this function is recorded along with the model parameters. Together, they propagate backward through the neural network architecture. In the next round of the training process, the algorithm updates and adjusts the parameters aiming to achieve a better performance. Finally, a well-trained model is generated with a parameter combination that produces an optimized evaluation function.

Fig.~\ref{fig:motivation} presents the training process of three deep learning models, Variational Auto-Encoders(VAE)~\cite{vae}, Bidirectional Recurrent Neural Networks(Bi-RNN)~\cite{birnn} and Gated Recurrent Unit(GRU)~\cite{gru}.
The Y-axle is the normalized values of evaluation functions (the lower the better) and the X-axle is the cumulative time.
As we can see from, considering the first 20\% of the total training time, 65\%, 98\%, and 90\% of the maximum reduction has been achieved for VAE, Bi-RNN and  GRU, respectively. 
Since they are trained individually on a physical machine, resources are fully occupied by each model without competitions. Therefore, given the same resource amount, the training gain and efficiency is decreasing as the time goes on.


\begin{figure}[ht]
\centering
         \begin{minipage}[t]{0.48\linewidth}
\centering
	\includegraphics[width=1\linewidth]{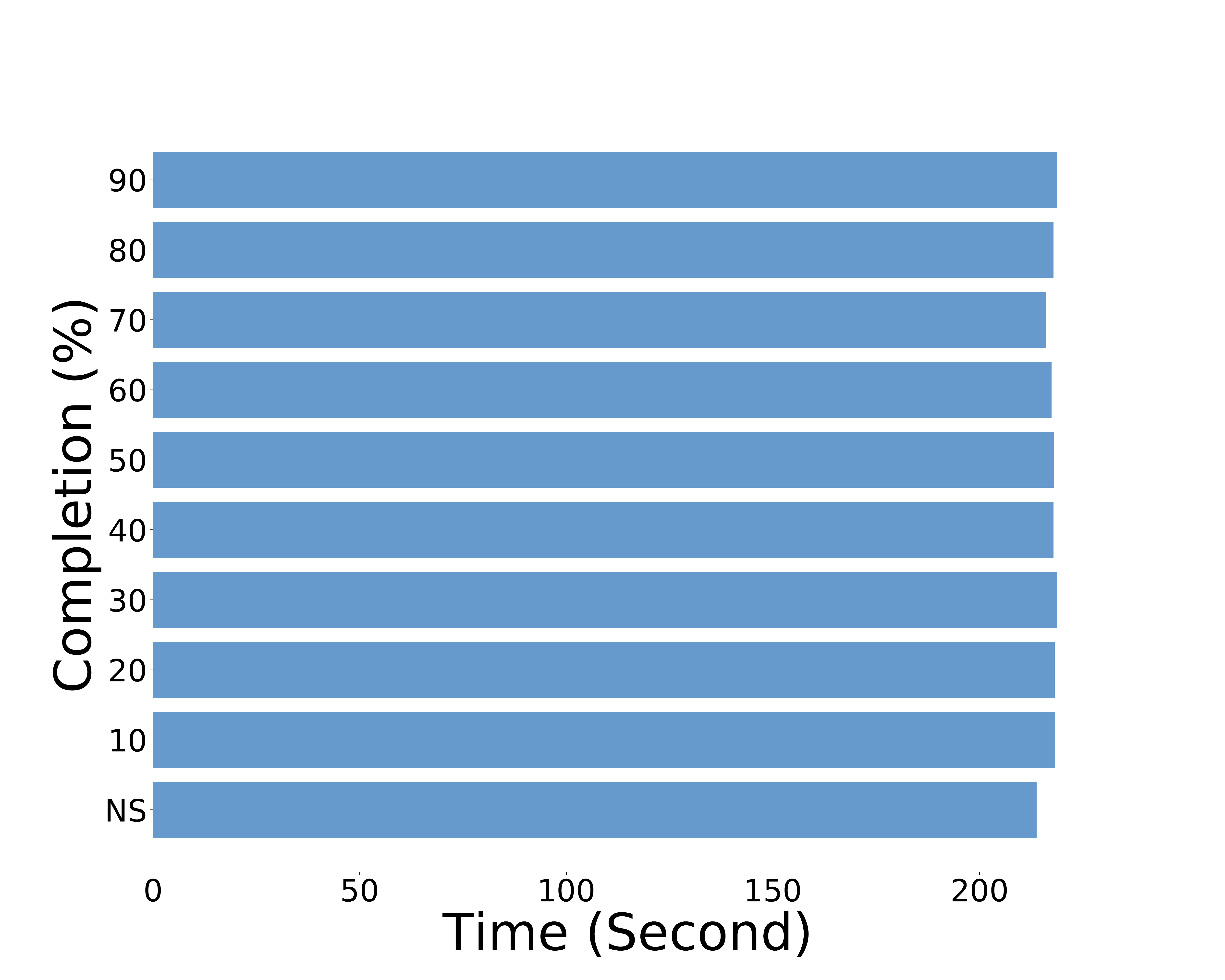}
	\caption{Overhead of saving and resuming VAE at different progress stage}
	\label{fig:vae}
      \end{minipage} 
      ~
      \begin{minipage}[t]{0.48\linewidth}
\centering
	\includegraphics[width=1\linewidth]{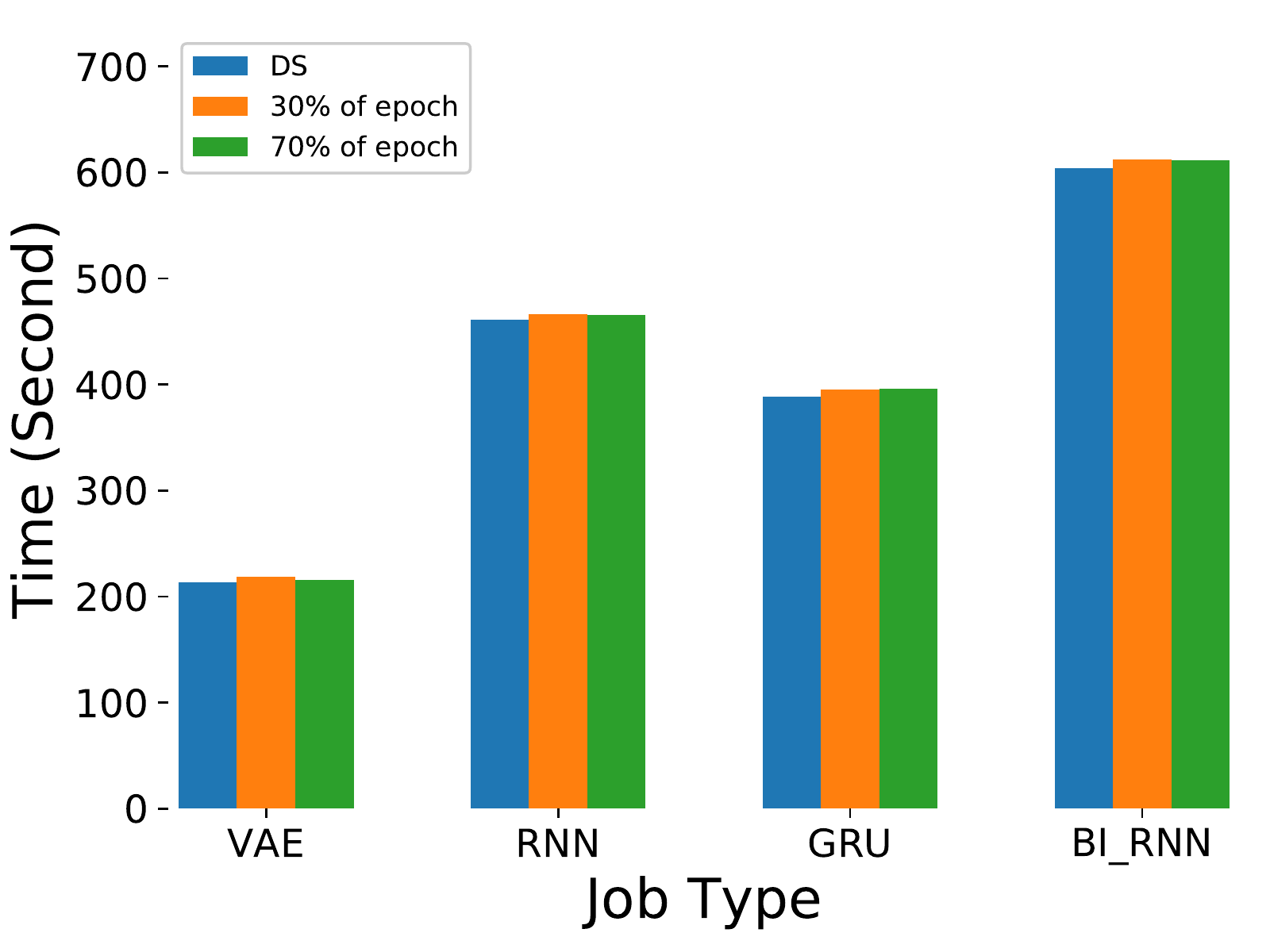}
	\caption{Overhead of saving and resuming deep learning models at 30\% and 70\%}
	\label{fig:learning}
      \label{fig:fixjob:ds}
      \end{minipage} %
\end{figure}

\subsection{Saving and Resuming a Container}

As analyzed above, a learning model, at any given iteration, contains the neural network design (the algorithm itself) and values of the network parameters that it has trained. That is to say the model progress can be saved during and after training. When necessary, the saved model can be resumed from exactly where they left off.

We experimentally study the overhead of saving and resuming learning models and compare to the nonstop training process.
The experiments are conducted under Tensorflow~\cite{tensorflow} and Pytorch~\cite{pytorch}, where the training jobs are running inside a Kubernetes container.
Fig.~\ref{fig:vae} shows the experiments of training a  VAE model, where we save the model at a given percentage according to the total number of iterations and resume it immediately after saving completes. 
Comparing to a nonstop training, denoted as $NS$ on the figure, 
we find that the overhead is minimum with all less than 5 seconds and consistent with a standard deviation value of 1.43 across different percentages during the train process. 

When saving a model in Tensorflow, there are four files generated internally. 

\begin{itemize}
\item Meta graph file: This is a protocol buffer which saves the complete Tensorflow graph.

\item Index file: It stores the list of variable names and shapes saved. 

\item Data file: This is the file which contains all the values of the weights, biases, gradients and all the other variables saved. 

\item Checkpoint file: It keeps a record of latest checkpoint files saved.

\end{itemize} 
 
When restoring the model, the system read these files and restart on where it stopped.

Next, we investigate the overhead with different deep learning models. Fig.~\ref{fig:learning} 
illustrates the comparison between nonstop training and the ones with saving/resuming.
The results of completion time are similar as shown on the figure that the overhead is minimum and consistent across different models. 




\section{SpeCon System Design}

\sol~ is implemented on top of Kubernetes, a dominant 
container-orchestration platform designed to automate the deployment, scaling, and management of containerized applications. 
In this section, we present the system architecture of \sol in details, including its key modules and their functionalities as well as design logics.

\subsection{Framework of Kubernetes}

Generally, a cluster of containers comprises managers and workers. 
Managers are responsible for system management, such as allocating the resources and handling the failures.
Workers are in charge of hosting the containers and executing the workloads.
In a Kubernetes cluster of containers, a pod is a group of one or more containers, which are hosted on the same worker and share the lifecycle as well as storage resources. 

Typically, there are  6 basic units in a Kubernetes cluster. 
The manager nodes consist of API Server, Controller Manager
and Default Scheduler and etcd. The Kuberlet and Service Proxy resident in workers. 

\begin{itemize}
\item API Server: It is the main management point of the entire
cluster and processes REST operations, validates them,
and updates the corresponding objects in storage.

\item Controller Manager: It runs controllers, which are the
background threads that handle routine tasks.

\item Default Scheduler: It watches for newly created Pods that have no
node assigned and is responsible for placement of pods
on workers.

\item etcd: It is a distributed data storage solution that stores
all the data, e.g. configuration, state, and metadata.

\item Kuberlet: It is responsible for maintaining a set of pods,
which are composed of one or more containers, on a local
system.

\item Service Proxy: It maintains network rules on nodes, e.g.
implementing a form of virtual IP for services.

\end{itemize}

\subsection{SpeCon System Architecture}

\begin{figure}[ht]
\vspace{-0.1in}
\centering
\includegraphics[width=0.8\linewidth]{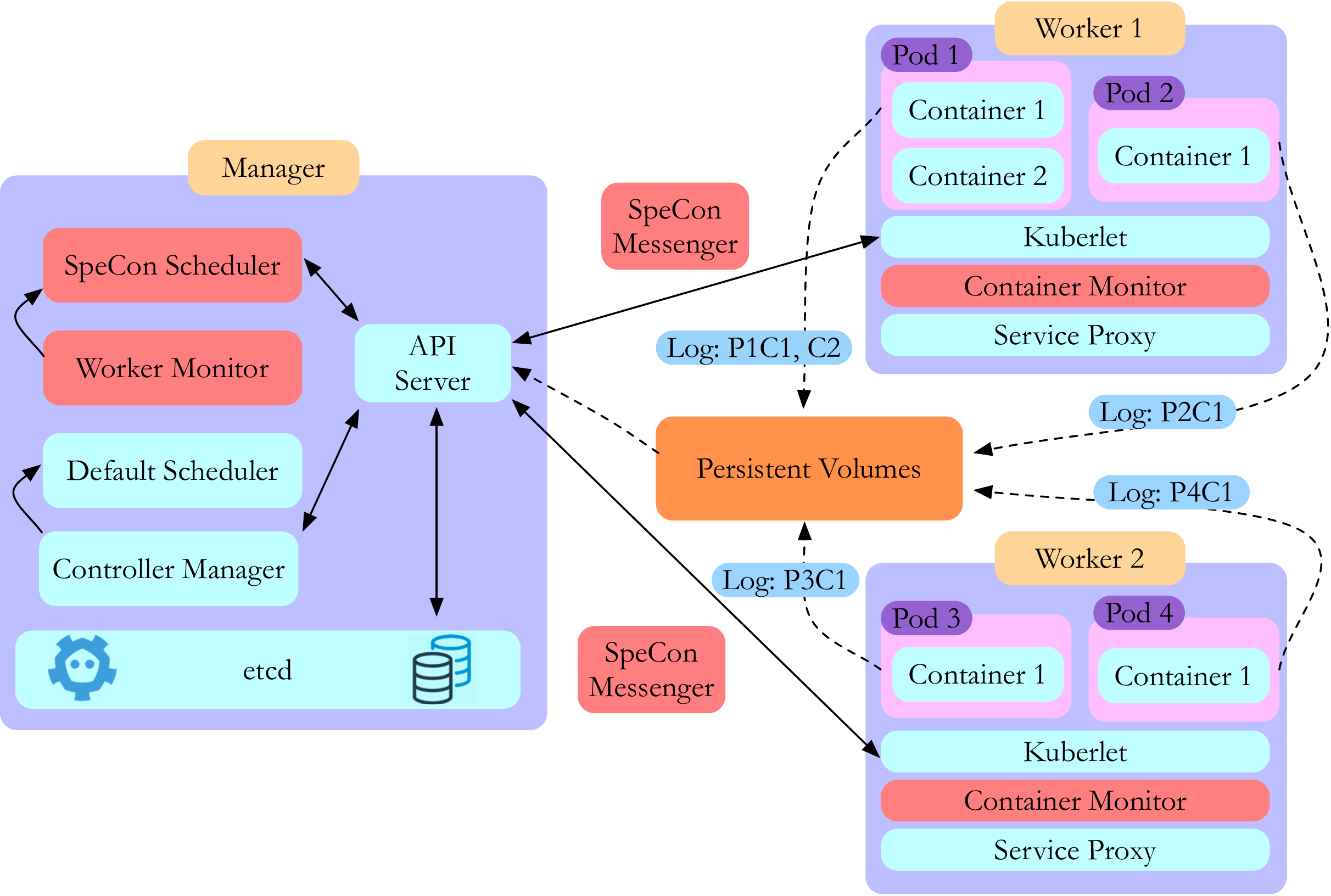}
\caption{SpeCon System Modules}
\label{fig:specon}
\vspace{-0.1in}
\end{figure}

Fig.~\ref{fig:specon} draws the \sol~ architecture in a 3-node cluster, which consists of 1 manager and 2 workers. 
Since \sol~ is built on top of Kubernetes, Fig.~\ref{fig:specon} includes the above mentioned 6 components in light blue.
Besides, it contains 4 key modules of \sol, {\bf Worker Monitor} and {\bf SpeCon Scheduler} on the manager side, 
{\bf Container Monitor} on the worker side, {\bf SpeCon Messager} that works in between. Particularly, 

\begin{itemize}

\item Container Monitor: It runs on the worker side, which provides the resources, such as CPU and memory, for containers to execute the training jobs. It keeps tracking the progress of each model from two perspectives, resource usages and current returns of their evaluation functions. Based on the collected data, it categorizes training jobs into different stages and active the corresponding algorithm (details in Section~\ref{sol}).

\item Worker Monitor: It resides on manager side, which is responsible for the overall system management. It monitors
all the workers in the cluster, such as the number of running jobs and the resource availabilities on each of them.
Moreover, it processes the messages from workers and communicates with \sol~ scheduler for further actions.  

\item SpeCon Scheduler: It works on the manager side. When the worker monitor decides that a training job needs to be migrated to another worker node, it utilizes the data that gathered by worker monitor and execute the algorithm to calculate a score for each of the worker, and then, select a most desirable node to host this migrated training job (details in Section~\ref{sol}).

\item SpeCon Messager: It generates heartbeat messages that transfer between managers and workers. For example,
workers, periodically, send a message to notify the manager such as the number of training job hosted on themselves and their categories.
Additionally, whenever a job needs to be migrated, workers make use of this channel to notify the manager immediately. 

\end{itemize}

The training jobs in \sol~ write their progress information into the log system in the system.  Particularly, \sol~
utilizes the Persistent Volume (PV~\cite{pv}), which is a piece of storage in the cluster and has a lifecycle independent of any individual Pod that uses the PV, to store the logs. This provides the stability to the \sol~ system.


\section{SpeCon Solution Design}
\label{sol}

\sol~ aims to improve the system performance by migrating the slow-growing
jobs to other workers to release resources for the fast-growing ones.
It achieves the improvement through three steps, (1) identify slow-growing jobs;
(2) select the most desirable host; (3) rebalance the workers. 
The Table~\ref{table:notation} summarizes the parameters and functions that
are used in \sol.

\begin{table}[ht]
	\centering
	\caption{Notations}
	\scalebox{0.95}{
		
		\begin{tabular}{ | c | c |}				
			\hline
			$w_i \in W$	      & The worker ID in the worker set  \\ \hline	
			$c_i \in C$ &	The container ID in the container set  \\ \hline	
			$E_{w_i, c_j}(t)$  & The value of evaluation function of $c_j$ on $w_i$ at time $t$  \\ \hline 
			$G_{w_i, c_j}(t)$   & The growth of evaluation function of $c_j$ on $w_i$ at time $t$ \\ \hline
			$R(w_i, t)$ & Resource consumption of $w_i$ at time $t$ \\ \hline
    		    $PC$ & The progressing category set \\ \hline
			$WC$ & The watching category set  \\ \hline
			$CC$ & The converged category set \\ \hline
			$\alpha$ & Categorization threshold \\ \hline
			$t_m - t_{m-1}$ & Categorization interval \\ \hline
			$T$ & Candidate set \\ \hline
			$S_{w_i}$ & The scoring set that stores weighted scores for each worker \\ \hline
			$bf$ & The balance factor for an uniform distribution \\ \hline
				\end{tabular}	
	}
	
	\label{table:notation}
\end{table}

\subsection{Identifying the slow-growing jobs}

\begin{algorithm}[!t]
\caption{Container Monitor: $c_{j}$ on $w_i$}
\label{alg:1}
\begin{algorithmic}[1]
\STATE System Parameters: 
						\par \quad  $\text{Categorization Interval}: t_m - t_{m-1}$ 
						\par \quad  $\text{Categorization Threshold}: \alpha$ 
						\par \quad  $\text{Evaluation Function}: E_{w_i, c_j}(t)$
						
\STATE Initial Settings: 
						\par \quad  $\forall c_{id} \in w_i, PC_{w_i}\text{.insert} (c_{id})$
						\par \quad  $\forall c_{id} \in w_i, c_{id.\text{migrated}} = \text{False}$
						
\item[]						

\FOR {$c_{j} \in w_i$ \& $c_{j.\text{migrated}} = \text{False}$}

	\STATE At time $t_m$: $\text{Calculate}~ E_{c_j, w_i}(t_m)$
	
	\STATE $G_{w_i, c_j}(t_m) = |E_{w_i, c_j}(t_m) - E_{w_i, c_j}(t_{m-1})|$

\IF {$G_{w_i, c_j}(t_m) < G_{w_i, c_j}(t_{m-1})$ \& $G{w_i, c_j} (t_m) < \alpha$}

	\IF {$c_j \in PC_{w_i}$ \& $c_j \notin WC_{w_i}$ \& $c_j \notin CC_{w_i}$}
		\STATE $PC_{w_i}\text{.remove}(c_j)$
		\STATE $WC_{w_i}\text{.insert}(c_j)$
	\ENDIF
	
	\IF {$c_j \notin PC_{w_i}$ \& $c_j \in WC_{w_i}$ \& $c_j \notin CC_{w_i}$}
		\STATE $PC_{w_i}\text{.remove}(c_j)$
		\STATE $WC_{w_i}\text{.insert}(c_j)$
	\ENDIF	

	\IF {$c_j \notin PC_{w_i}$ \& $c_j \notin WC_{w_i}$ \& $c_j \in CC_{w_i}$}
		\STATE $c_j \text{remains in}~ CC_{w_i}$
				\IF {$|PC| + |WC| > 1$}
				\STATE $\text{SpeCon-Messager}(w_i, c_j).\text{Request-Reallocation}$
				\ENDIF	
	\ENDIF

\ELSIF {$G_{w_i, c_j}(t_m) \geq G_{w_i, c_j}(t_{m-1})$ \& $G{w_i, c_j} (t_m) < \alpha$}
	\STATE $c_j$ stays in the current category
	
\ELSIF {$G{w_i, c_j} (t_m) > \alpha$}
		\STATE $PC_{w_i}\text{.insert}(c_j)$
		\STATE $WC_{w_i}\text{.remove}(c_j)$
		\STATE $CC_{w_i}\text{.remove}(c_j)$
		
\ENDIF

\ENDFOR

\end{algorithmic}
\label{alg:worker}
\end{algorithm}

In \sol, we consider a cluster of containers that hosted on workers. Inside the cluster, we have managers (denoted as $M$). In addition, we denote $c_i \in C$ to be  the set of containers and
$w_i \in W$ to be the set of workers. Therefore, the relation, $c_m \in w_n$, represents that a container, $c_m$, is running on 
a worker $w_n$. 

In our problem setting, training jobs are running inside containers, where each container hosts one specific job. 
Consequently, each container, $c_i$, can be resolved as one particular training job in our setting. As analyzed in the previous sections, each job has a predefined evaluation function. During the whole training process, it is a time series function, 
where at the end of each iteration the return value updates. When querying in the middle of an iteration, the previous value will be returned. Based on the query time, Equation~\ref{eq:evaluation} presents the growth of the training job in a given interval, $t_2 - t_1$.

\begin{equation}
\label{eq:evaluation}
	G_{w_i, c_j}(t_2) = E_{w_i, c_j}(t_2) - E_{w_i, c_j}(t_1)
\end{equation}
, where $c_j$ is a container that runs a training job on $w_i$ with 
 $E_{w_i, c_j}(t)$ as the evaluation function.

According to the values of  $G_{w_i, c_j}(t)$ , \sol~ classifies $c_i$ into 3 different categories. 
\begin{itemize}
\item Progressing Containers ($PC$): The jobs inside these containers are still in the fast growing stage such that their evaluation function is progressing quickly.

\item Watching Containers ($WC$): When a training job join $WC$ category, it indicates that this job is slowing down on the gain of the evaluation function. In this zone, the gain starts jittering and, depending on the model itself, it may speedup the progress again.

\item Converged Containers ($CC$): If training jobs in $WC$ continue slowing down, they are inserted into a Converged Container category, which suggests that they are slow-growing on their evaluation functions and unlikely to speedup again.

\end{itemize}

Based on the above analysis, we develop the Algorithm~\ref{alg:worker} in \sol~ to keep tracking the progress of training jobs on each worker. Please note that, in order to avoid the unnecessary network traffic and maintain scalability, the containers in \sol~ will be reallocated at most once. 

As shown in Line 1, the Algorithm~\ref{alg:worker} on each worker first fetches key parameters from the system. The categorization interval ($t_m - t_{m-1}$) determines how frequent a worker checks containers' log files, which store return values of evaluation functions. The categorization threshold, $\alpha$, is a percentage value that \sol~ uses to decide 
whether the progress is growing quickly. Then, in Line 2, the algorithm initializes parameters such that all containers are inserted into $PC$ when they join a worker and their migration indicator is set to false at the beginning.

At time $t_m$, it reads the current value of  the evaluation function and calculates the gain during the previous categorization interval, $t_m - t_{m-1}$ (Line 3-5). If the gain is less than the previous round and the threshold $\alpha$, \sol~ updates the category of this container based on the following condition (Line 6-16).
\begin{itemize}
\item If the container is currently in the category $PC$, but not $WC$ and $CC$, it will be  removed from $PC$ and inserted to $WC$. It means that this job starts slowing down, but not stable yet.

\item If the container is currently in the category $WC$, but not $PC$ and $CC$, it will be removed from $WC$ and inserted to $CC$. It indicates that this job has started its convergence process.

\item If the container is currently in the category $CC$, but not $PC$ and $WC$, it suggests that the value of the gain during the interval continues decreasing and it has converged. In this condition, it will remain in the $CC$ category and the worker will call \sol~ messenger to send a migrated request of $(c_j, w_i)$ to the manager.
\end{itemize}

In a scenario that the current gain is less than $\alpha$, but larger than the previous gain value, this container stays in the current category. This is due to the fact that when a training job goes on, the model randomly selects and updates parameters at each iteration that leads to an unstable trends (bouncing growth values). In this stage, the container remains in the same category and waiting for next round (Line 17-18). 

At the moment that the gain becomes larger than the threshold, we reset the container's category to $PC$, which is utilized to accommodate a sudden change of the evaluation function and possible errors from the previous rounds (Line 19-22).

\begin{algorithm}[!t]
\caption{Speculative Scheduling of $c_j$ in $w_i$ on Manager}
\label{alg:2}
\begin{algorithmic}[1]

\STATE Parameter Initialization: 
		\par \quad $\text{Candidate\_Set}: T  = \phi$
		\par \quad $\text{Weights}: w_{pc}, w_{wc}, w_{cc}$
		\par \quad $\text{Resource Consumption on} ~w_i ~\text{at time} ~t: R(w_i, t)$
\item[]

\FOR {$w_j \in W$}
	\STATE $ S_{w_j} = |PC_{w_j}| \times w_{pc} + |WC_{w_j}| \times w_{wc} + |CC_{w_j}| \times w_{cc}$
\ENDFOR

\FOR {$w_j \in W$}
	
	\STATE Find Min($S$)
	
	\IF {$S_{w_j} =$ Min$(S_{w_j \in W})$}
		\STATE $T\text{.insert}(w_j)$
	\ENDIF	

\ENDFOR	

\IF {$w_i \in T$ }
	\STATE $c_{id.\text{migrated}} = \text{True}$
	\STATE $c_j$ remains on $w_i$
	
\ELSE
	\FOR {$w_j \in T$}
		\IF {$|T| = 1$}
			\STATE Save $c_{j}$ on $w_i$
			\STATE $c_{id.\text{migrated}} = \text{True}$
			\STATE Restore $c_j$ on $w_j$;				
		\ELSIF {$|T| > 1$}
			\STATE Rank $w_j \in T$ with $R(w_j, t)$
			\STATE Save $c_{j}$
			\STATE $c_{id.\text{migrated}} = \text{True}$
			\STATE Restore $c_j$ on $w_j$ with lowest $R(w_j, t)$;
		\ENDIF	
	\ENDFOR
	
\ENDIF			

\end{algorithmic}
\label{alg:manager}
\end{algorithm}

\subsection{Scheduling on slow moving jobs}
The container monitor that runs each worker keeps tracking the evaluation functions and collects data that stores in the persistent volume. Whenever receives a request from workers, the manager responses the reallocation message, e.g. $(w_i, c_j)$, by executing the Algorithm~\ref{alg:manager}. The main objective of Algorithm~\ref{alg:manager} is to select the most desirable worker node to host this container, e.g. $c_j$. 

\sol~ utilizes a weighted scoring algorithm to rank worker nodes. As shown in Line 1 of Algorithm~\ref{alg:manager}, it first initializes parameters, which include a candidate set ($T$) that uses to store targeted workers and it is set to empty initially. Additionally, \sol~ maintains predefined weights for each category as well as resource consumption functions. 

For each worker node in the cluster, it calculates a score based on the number of containers in each category and its weights. \sol~ aims to improve the efficiency by allocating more resources to fast-growing jobs and limit resources for slow-growing ones. With this objective in mind, the values of weights have a relationship of $w_{pc} > w_{wc} > w_{cc}$, where priorities are given to progressing containers (Line 2 - 3).

Given the scores of the workers, the algorithm finds out workers that have the minimum score. Those workers are inserted into the candidate set. If multiple workers have the minimum score, it results in more than one candidates (Line 4-7). 
The algorithm checks whether $w_i$, the current host of $c_j$,  is in the set or not. Then the following logic will be executed. 

\begin{itemize}
\item If $w_i$ in candidate set,  worker $w_i$ will be returned, which means that container $c_j$ should continue running on $w_i$. In this case, the algorithm marks the container as migrated (Line 8-10). This manner reduces unnecessary overhead caused by migration. 

\item If $w_i$ is not in the candidate set, the algorithm takes two branches. (1) If there is only 1 worker in $T$, 
\sol~ select this worker for migration. (2) If $|T|>1$, \sol~ ranks candidates with their resource usages and selects
the one with least resource usage as the new host (Line 11-21). 


\end{itemize}

\subsection{Rebalancing workloads in the cluster}
Together with Algorithm \ref{alg:1} and \ref{alg:2}, \sol~ distributes the workload based on real-time returns from evaluation functions of containers
as well as current resource consumption on workers. 
At this stage, however, it could make inaccurate decisions due to missing information of finishing time.
For instance, if $\forall c_i \in W, c_{i.\text{migrated}} = \text{True}$ that means all jobs have converged,
therefore, the workload distribution in the cluster is fixed, which would result in an imbalanced cluster.
As intuitive example, when all 4 jobs become converged in a 2-node cluster, $W_1$ hosts $c_1$ and $c_2$ and
$W_2$ contains $c_3$ and $c_4$. In a scenario that $c_1$ and $c_2$ completes before $c_3$ and $c_4$, $W_1$ runs
without any workloads. 

In a real cloud environment, it is challenge and costly to obtain an accurate finish time due to the various implementations, dynamic workloads and resource competition. In our solution, \sol~ incorporates the converged duration, which is time length between when a container is marked as converged and current time to compare active containers and re-distribute the workload in a cluster by using Algorithm \ref{alg:3}.

The algorithm prepares the required parameters in Line 1. Then, it calculates the
sum of active jobs on workers 
across the cluster and computes the converged duration for each container $c_j$ on each worker $w_i$.
The each converged duration, $d_{c_j}$, is inserted into a set $D$ (Line 2-6).
If there is no active jobs on a particular worker $w_i$, its id would be inserted to the candidate set $T$,
which stores the workers that can take more workload (Line 7-8).
Given $tj$ and $|W|$, the balanced factor, $bf$, is obtained in Line 9. The value of $bf$ is based on uniform distribution.

Depending on the number of workers in the candidate set $T$, Algorithm \ref{alg:3} takes the following two branches. 

\begin{itemize}
\item If $T$ is nonempty, it suggests that, at least, one worker is idling. 
For each idling worker $w_i$ in $T$, 
the algorithm assigns $c_j$, which has the smallest $d_{c_j}$ to $w_i$ if the number of active containers on it 
is less than $bf$ and marks $c_j$ to be rebalanced (Line 10-18).

\item If $T$ is empty, it means that every worker runs some active containers.
Then, \sol~ enumerates all workers and finds the ones that host less jobs than $bf - 1$, which indicates they
can hold, at least, one more job. These workers are inserted into $T$ (Line 19-22).
Then, the algorithm finds the $c_j$ with the smallest $d_{c_j}$ and, in the meanwhile, $c_j$ runs on a worker $w_j$ such that
$w_j \notin T$ (Line 23-25). It basically avoids the scenario that $c_j$ assigns to its current host.
With a nonempty $T$, \sol~ assigns the previously found $c_j$ to $w_i$ which has room for an additional container and
marks this container to be rebalanced (Line 26-31). 

\end{itemize}

\begin{algorithm}[!t]
\caption{Rebalance Active Container in the Cluster}
\begin{algorithmic}[1]

\STATE Parameter Initialization: 
		\par \quad $\text{Candidate\_Set}: T  = \phi$
		\par \quad $\text{Resource Consumption on} ~w_i ~\text{at time} ~t$: $R(w_i, t)$
		\par \quad $|w_i|$: The number of active containers on $w_i$
		\par \quad $t_{c_j}$: The timestamp when $c_j$ is marked as migrated
		\par \quad $t, D$: The current timestamp, duration set $D$
		\par \quad $bf, |W|$: The balance factor, total number of workers
		
\item[]
		
\FOR {$w_i \in W$}

	\STATE $sum = sum + |w_i|$	

	\FOR {$c_j \in w_i$}
		\STATE $d_{c_j} = t - t_{c_j}$
		\STATE $D$.insert($d_{c_j}$)
	\ENDFOR	 
			
	\IF {$|w_i| = 0$}	
	\STATE $T$.insert($w_i$)
	\ENDIF
	
\ENDFOR	

	\STATE $bf = \lfloor sum \div |W| \rfloor $

\IF {$|T|>0$}
	\FOR {$w_i \in T$}
		\IF {$|w_i| < bf$}
			\STATE Find $c_j$ with  Min($d_{c_j}$) in $D$	
			\STATE $D$.remove($d_{c_j}$)
			\STATE Migrate $c_j$ to $w_i$
			\STATE $c_j$.rebalanced = True
		\ELSE
			\STATE Continue;
		\ENDIF
	\ENDFOR

\ELSIF {$|T| = 0$}
	\FOR {$w_i \in W$}
		\IF {$|w_i|  < bf - 1$}
			\STATE $T$.insert($w_i$)
		\ENDIF	
	\ENDFOR	

	\FOR {$w_i \notin T$}	
	 	\STATE $\forall c_j \in w_i$
		\STATE ~~~~Find $c_j$ with  Min($d_{c_j}$) in $D$
			\FOR {$w_j \in T$}
				\IF {$|w_i| < bf - 1$}
					\STATE Migrate $c_j$ to $w_j$
					\STATE $c_j$.rebalanced = True	
				\ELSE
					\STATE Continue;				
				\ENDIF
			\ENDFOR
	\ENDFOR				
\ENDIF		

\end{algorithmic}
\label{alg:3}
\end{algorithm}


\section{Evaluation}

In this section, we evaluate the performance of
\sol~ through intensive cloud-executed experiments.

\subsection{Experimental Setup}
\subsubsection{Implementation, Testbed and Workloads}

\sol~ is integrated into Kubernetes 1.17 as plug-in modules that reside on both managers and workers. 
It receives tasks from the manager, 
directs the given tasks to workers, monitors the evaluation functions and balances the workloads.

We build a testbed on a NSF Cloudlab~\cite{cloudlab} datacenter that hosts at University of Utah.
Specifically, we use multiple M510 as our physical machines that contain two 
8-core Intel Xeon D-1548 at 2.0 GHz, 64GB ECC Memory,
and 256 GB NVMe flash storage.
Two clusters have been built to evaluate \sol:
{\bf Cluster 1}: 1 Manager and 4 Workers
{\bf Cluster 2}: 1 Manager and 8 Workers

\sol~focuses on accelerating the training process of deep learning models. We evaluate it with two 
popular open-source platforms, Pytorch (P)~\cite{pytorch} and Tensorflow (T)~\cite{tensorflow}. 
As shown in  Table~\ref{table:workload}, we choose 5 different model on those two platforms as our workloads that execute inside
containers.

\begin{table}[ht]
	\centering
	\caption{Tested Deep Learning Models}
	\scalebox{0.95}{
		
		\begin{tabular}{ | c | c | c | }				
			\hline
			Model	      & Loss Function & Platform  \\ \hline	
			Variational Autoencoders (VAE) &	Recon. Loss	 & P/T  \\ \hline	
			Gated Recurrent Unit (GRU)  & Quadratic Loss & P/T \\ \hline 
			Bidirectional-RNN   & Softmax & P/T \\ \hline
    		    Recurrent Network (RNN) & Softmax & P/T \\ \hline
			Dynamic RNN & Softmax & P/T \\ \hline
		\end{tabular}	
	}
	
	\label{table:workload}
\end{table}

\subsubsection{Evaluation Metrics}

The training processes of deep learning applications are computational intensive.
They are more sensitive to CPU powers than memory spaces and network bandwidth.
The following evaluation metrics are considered in our experiments.
\begin{itemize}
\item {\bf Completion time}: The completion time of each individual job in the cluster.
\item {\bf Average completion time}: The average completion time of among all the workload in the cluster.
\item {\bf Makespan}: The makespan is the time between the initialization of the first job and the completion of the last one across all workers.
\end{itemize}

When multiple models are training on the same worker, the jobs create a highly dynamic computing environment in terms of resource competition.
Therefore, we design the following submission schedules to ensure a comprehensive evaluation.
\begin{itemize}
\item  {\bf Fixed Schedule}: The time to
launch containers follows a fixed interval. It simulates an administrator controlled cloud environment.

\item {\bf Random Schedule}: The
time to launch a job is randomly selected within an interval.
It simulates an user-specified, first-come-first-serve cloud environment.

\end{itemize}

As for the weight of each category that utilized by Algorithm~\ref{alg:2}, we use $w_{pc} = 2, w_{wc} = 1.5$ and 
$w_{cc} = 1$. Due to the page limit, we omit the analysis of weights in this paper.  
Additionally, we compare \sol~ with the default scheduler in Kubernetes, which noted as {\bf DS} in the rest of the evaluation section. For a fair comparison, we utilize a fixed initial container placement, where workload are evenly distributed
based on the number of containers,
 for both $DS$ and \sol so that
only the algorithms in \sol~ can affect the performance.

\subsection{Fixed schedule}
\label{fixed}
Firstly, we conduct experiments with a fixed schedule such that containers with the same model image are submitted 
to the system one by one with a fixed interval. In this experiment, we use VAE model and 50 second interval.
Furthermore, we set $\alpha = 0.01, |t_m - t_{m-1}| = 30$ for \sol.

\begin{figure}[ht]
\centering
\includegraphics[width=\linewidth]{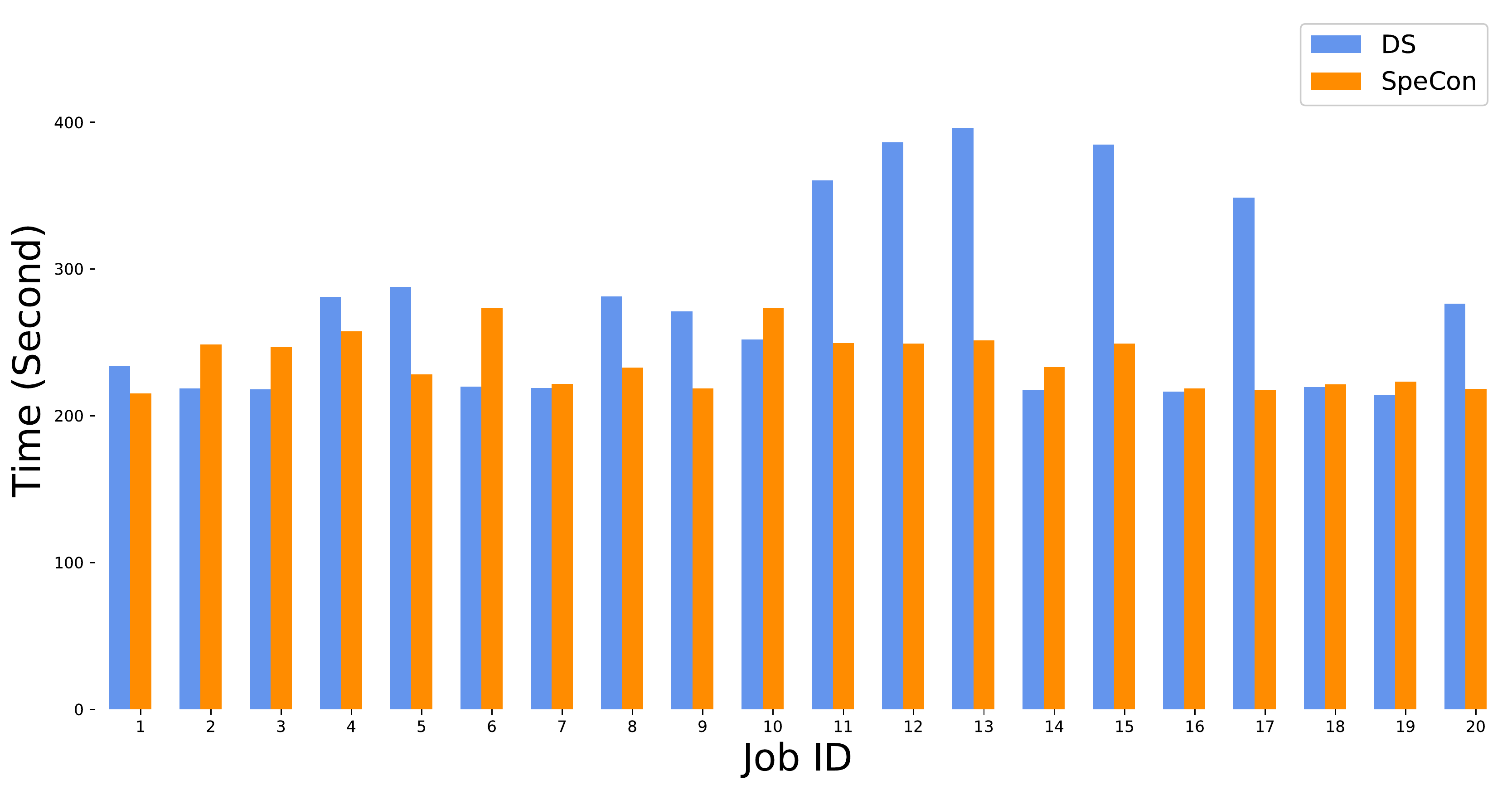}
\caption{Fixed Schedule with 20 VAE jobs and 50s interval ($\alpha = 0.01, |t_m - t_{m-1}| = 30$)}
\label{fig:20vae} 
\end{figure}

As shown on Fig.~\ref{fig:20vae}, we can clearly see that \sol~ outperforms $DS$ in terms of completion time.
The largest gain is found on Job-17, which reduces from 348.7s to 217.6s, that is a 37.6\% reduction.
There are 11 jobs out of 20 get a reduction the completion time. The average reduction in those 11 jobs is 83.6s  
(26.7\%). For those 9 jobs that fail to obtain improvement, the average increase is  18.3s (8.1\%).
Overall, the average completion time of \sol~ reduces from 275.2s to 234.7s (14.7\%).
With speculative scheduling on slow-growing containers, \sol~ is able to migrate them to other workers and release resources for
fast-growing jobs, which leads to reduction on completion time.
As for the makespan of the system, \sol~ reduces the value from 1260s to 1190s.

\subsection{Random schedule}
\label{random}
\subsubsection{Single model}


There are two key parameters used by \sol, $\alpha$ that determines the category of each job and
$|t_m - t_{m-1}|$ that decides when to start categorization.
We conduct experiments
to study the performance of \sol~ under various
parameter combinations.
In these experiments, we randomly submit jobs (single VAE model) within the interval [0, 300] to the cluster.


Table~\ref{table:parameters} presents the results when comparing to $DS$ and $\alpha = 0.05$, $|t_m - t_{m-1}| = 20, 25, 30$
as well as $|t_m - t_{m-1}| = 30$, $\alpha = 0.01, 0.05, 0.1$.
The data includes, the percentage of jobs that achieve reduction on completion time (Reduced),
the overall average completion time improvement among all the jobs (Overall), the reduction of the best job (Best), and
the improvement of makespan of the whole system (Makespan). The results demonstrate that \sol~
clearly outperforms $DS$ from different perspectives.
When comparing to the fixed schedule, degrees of improvements that achieved by \sol, with random submissions, reduced. 
The reason lies in the fact that random submissions creates challenges 
categorizing the running containers since there is delay
when we monitoring the evaluation function and the single model 
generates difficulties to select most desirable worker since the same training pattern has a higher 
probability leads to the same score for workers.

\begin{table}[ht]
	\centering
	\caption{Parameters in \sol}
	\scalebox{0.95}{
		
		\begin{tabular}{ | c | c | c | c | c |}				
			\hline
			Parameters	      & Reduced & Overall & Best & Makespan \\ \hline	
			0.05, 20s &	60.0\% & 5.7\% & 29.1\% & 6.1\% \\ \hline	
			0.05, 25s & 60.0\% & 6.8\% & 27.8\% & 7.5\% \\ \hline 
			0.05, 30s & 55.0\% & 3.2\% & 21.3\% & 4.0\% \\ \hline
			\hline
    		    0.01, 30s & 65.0\% & 9.6\% & 31.0\% & 8.2\% \\ \hline
			0.05, 30s & 55.0\% & 3.2\% & 21.3\% & 4.0\% \\ \hline    		    
			0.10, 30s & 60.0\% & 6.4\% & 23.9\% & 3.1\% \\ \hline
		\end{tabular}	
	}	
	\label{table:parameters}
	\vspace{-0.1in}
\end{table}




{\bf Remarks}: 
Based on the experimental results and analysis in Section~\ref{fixed} and \ref{random}, 
we can conclude that \sol~ can improve both the
system makespan and the completion time of individual learning
jobs with different parameter settings, compared to the original
Kubernetes system.
In the meantime, different parameter settings do affect the degree of improvement. 
Since values of $\alpha$ and $t_m - t_{m-1}$  indicate how quickly a given container enters $CC$ category in Algorithm 1,
they controls how many containers can be considered for migration and frequency of Algorithm 2. 
Therefore, the values are proportional to overhead, e.g. calculating the data for algorithms and
cost for migrating a container, that \sol~ brings to the system.
Consequentially, as the frequency of running Algorithm 1 and 2 decreases, 
the rooms scheduling and adjusting the system will be reduced. 
Therefore, the best $\alpha$ setting depends on the number
of active containers in the system, the deep learning
model in each container and the corresponding datasets.
Due to the page limit, we omit the theoretical analysis of parameters and, in the rest of the evaluation, the experiments utilize $|t_m - t_{m-1}| = 30, \alpha = 0.01$.



\subsubsection{Multiple models}

\begin{figure}[ht]
\centering
\includegraphics[width=\linewidth]{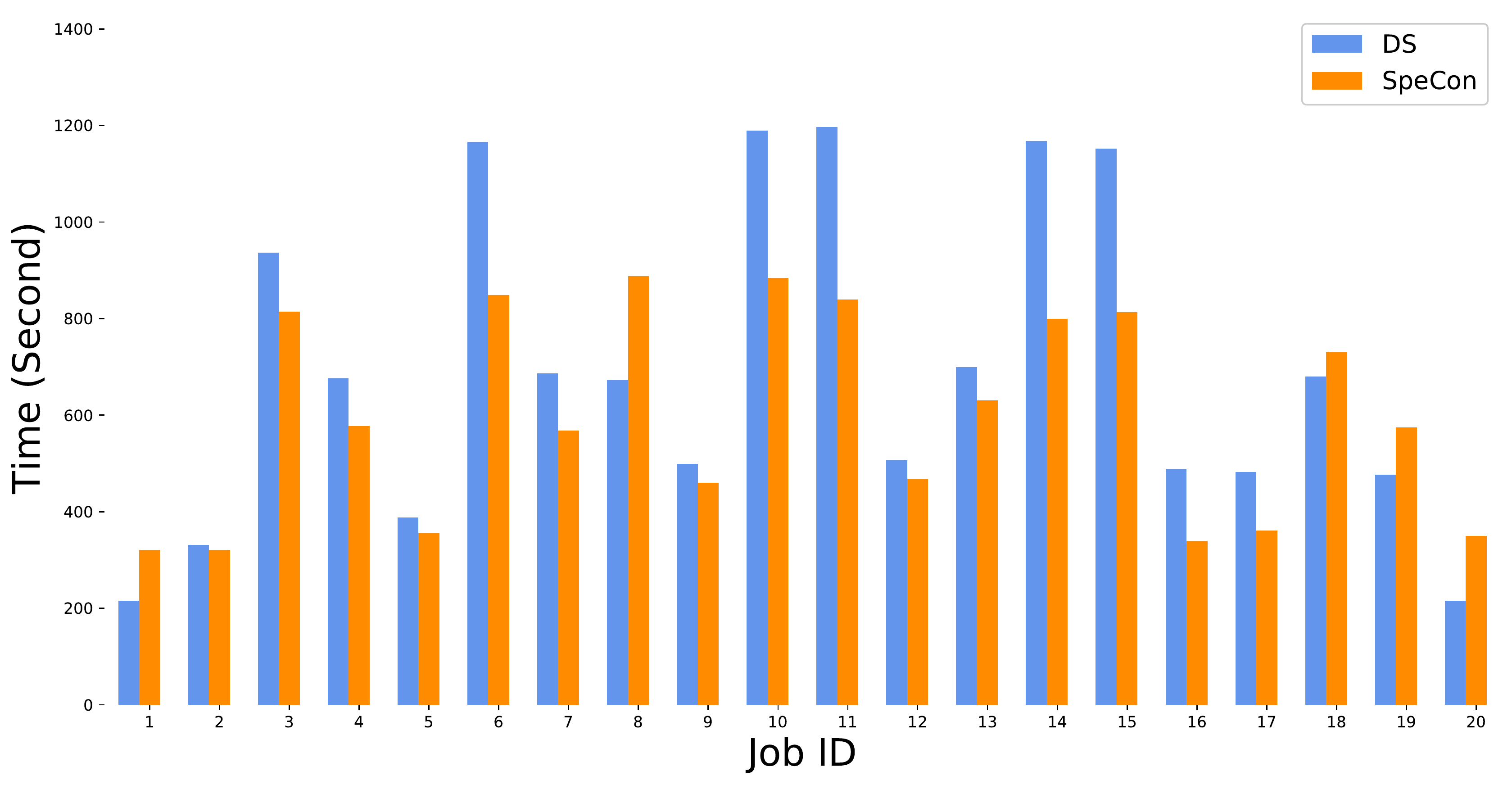}
\caption{Random schedule [0, 300] with 20 random jobs}
\label{fig:random20} 
\end{figure}

Next, we conduct experiments that employ a random selection of both models and platforms in Table~\ref{table:workload}.
Fig.~\ref{fig:random20} plots the same cluster with 20 randomly selected jobs submitted to the system within [0, 300].
Overall, there are 15 out of 20 jobs (75\%) record a reduced completion time.
The largest gain is obtained on Job-14, whose completion time reduced from 1167.8s to 799.2s (31.6\%).
The average improvement of those 15 jobs is 165.7s. For the other 5 jobs, the average increase is 110.8s.
The increase is cause by migrating slow growing containers to the same worker lead to more resource competition.
\sol~ sacrifices a small portion of jobs to improve others significantly.
With a system-wide consideration, \sol~ achieves 13.6\% reduction on the average completion time.
In terms of makespan in the cluster, \sol~ achieves 24.7\% reduction from 1421s to 1070s.

\begin{figure}[ht]
\centering
\includegraphics[width=\linewidth]{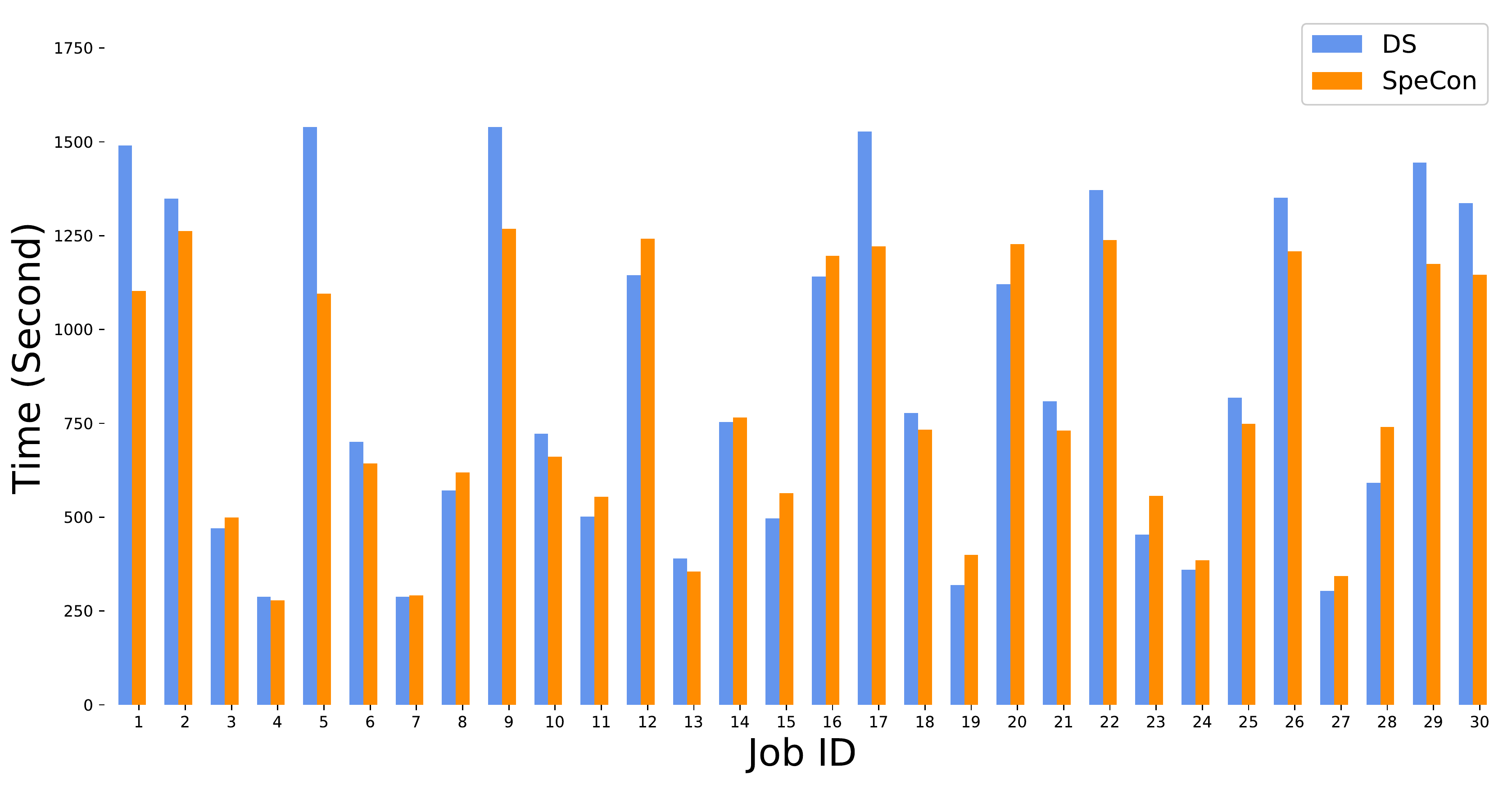}
\caption{Random schedule [0, 600] with 30 random jobs}
\label{fig:random30} 
\end{figure}

The previous experiments contain only single model (VAE). 
Next, we conduct experiments that employ a random selection of both models and platforms in Table~\ref{table:workload}.
In this experiment, we randomly launch 30 jobs within the interval of 
[0, 600]. 
Fig.~\ref{fig:random30} shows the completion time of these jobs.
Comparing with $DS$, 18 out of 30 jobs (60.0\%) finish training faster with \sol.
For those 18 jobs, the average improvement is 15.1\% and for reminder 12 jobs, the average 
increase is 10.6\%. 
Among the workloads, Job-5 gains the largest, 28.8\%, reduction from 1539.6s to 1095.5s (detailed explanations below). 
Overall, the average completion time improves from 865.7s to 800.2s, 7.6\% and makespan reduces from 1760s to 1498s, 14.8\%.

\begin{figure*}[ht]
   \centering
         \begin{subfigure}[b]{0.23\textwidth}
\centering
         \includegraphics[width=\textwidth]{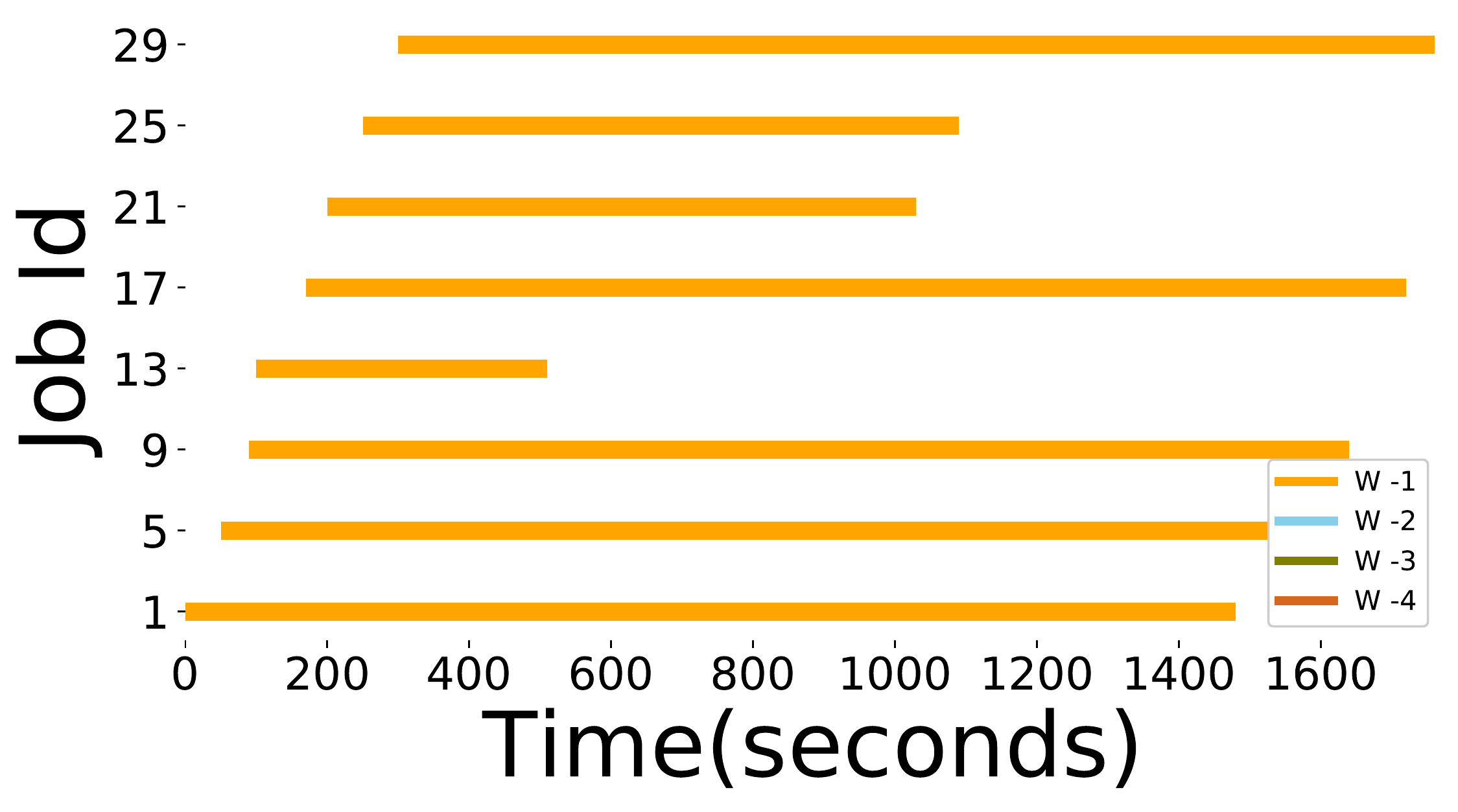}
\caption{Worker-1}
      \label{fig:random30:ds:node1}
      \end{subfigure} 
      \begin{subfigure}[b]{0.23\textwidth}
\centering
         \includegraphics[width=\textwidth]{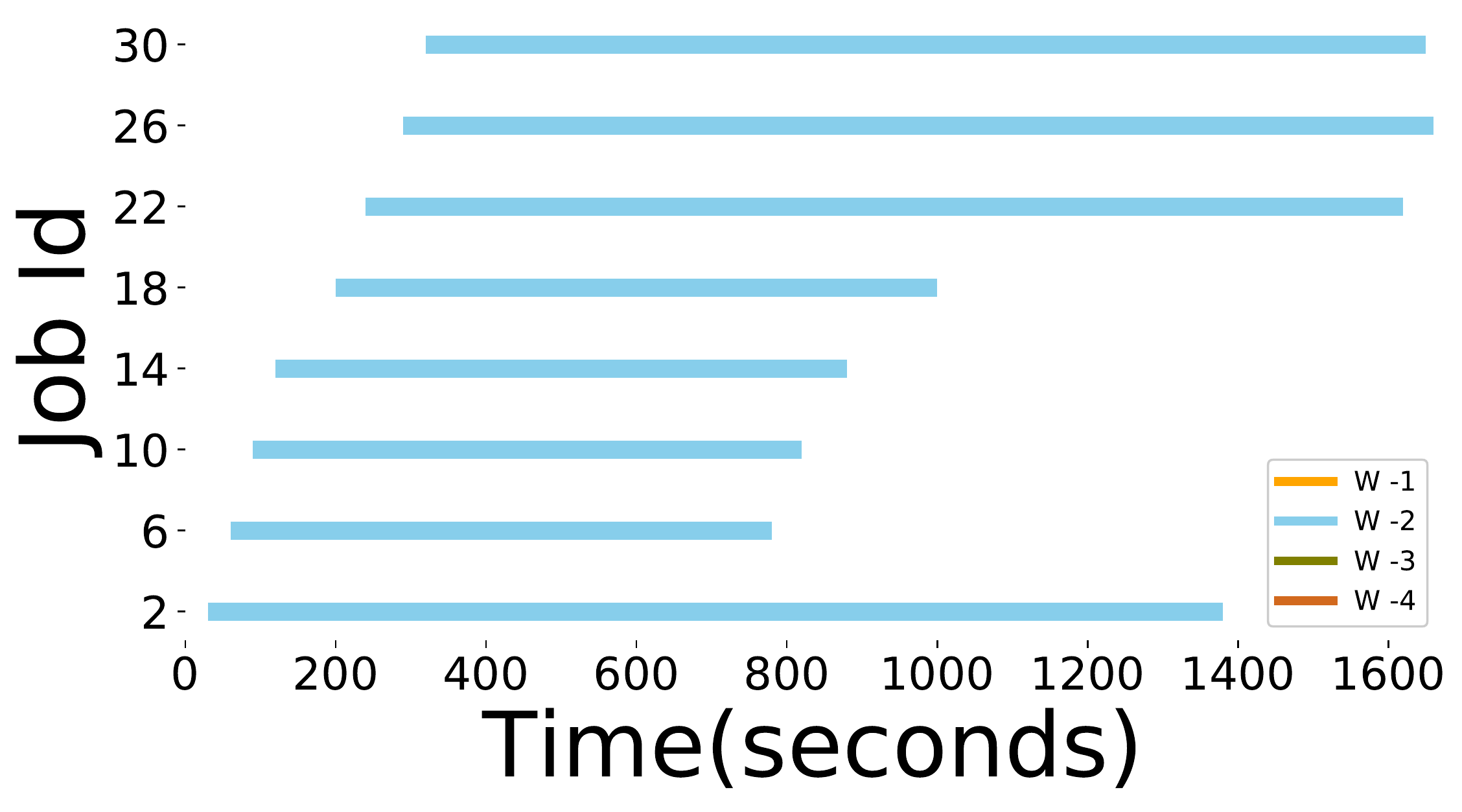}
\caption{Worker-2}
      \label{fig:random30:ds:node2}
      \end{subfigure} %
      \begin{subfigure}[b]{0.23\textwidth}
\centering
         \includegraphics[width=\textwidth]{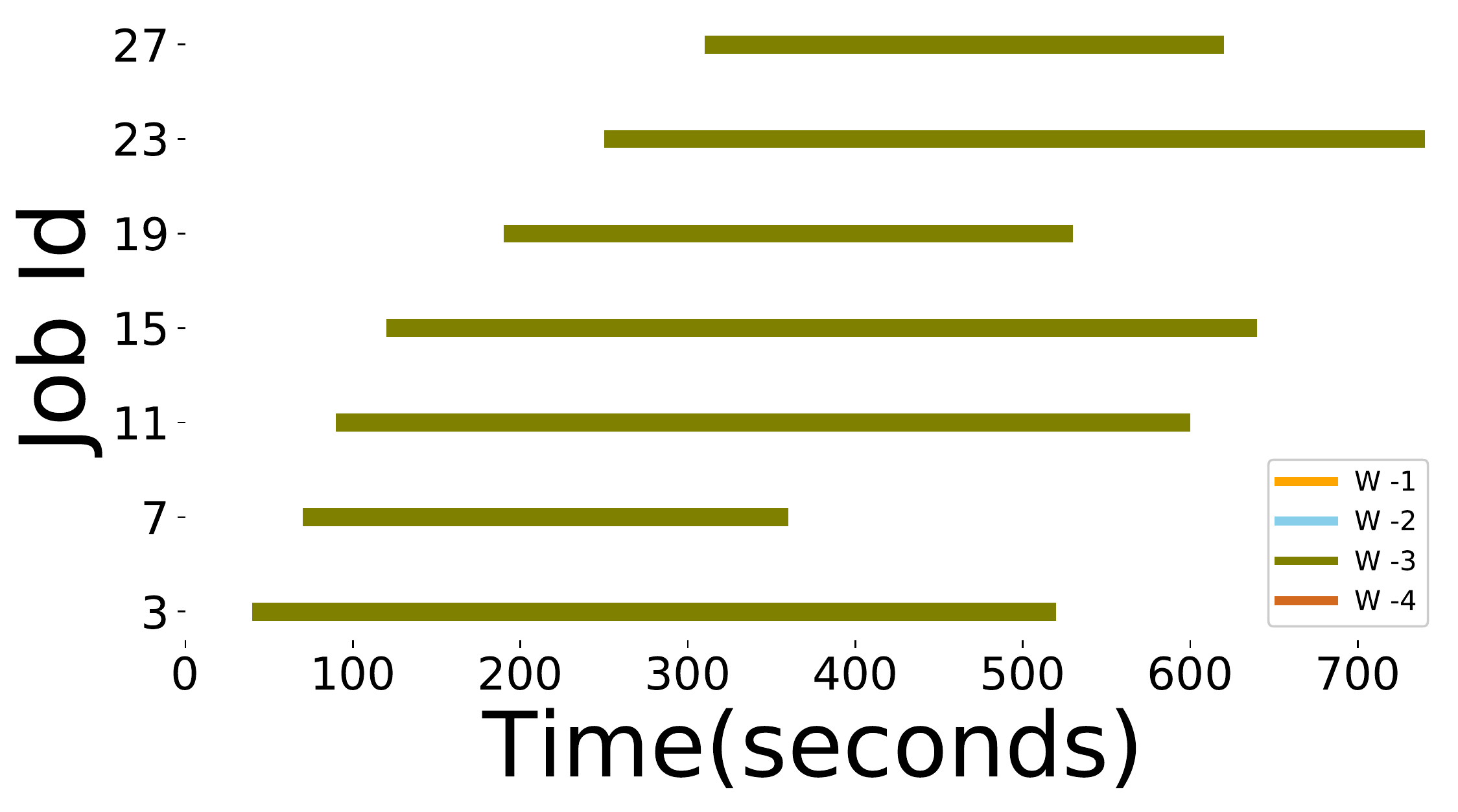}
\caption{Worker-3}
      \label{fig:random30:ds:node3}
      \end{subfigure} %
      \begin{subfigure}[b]{0.23\textwidth}
	\centering
         \includegraphics[width=\textwidth]{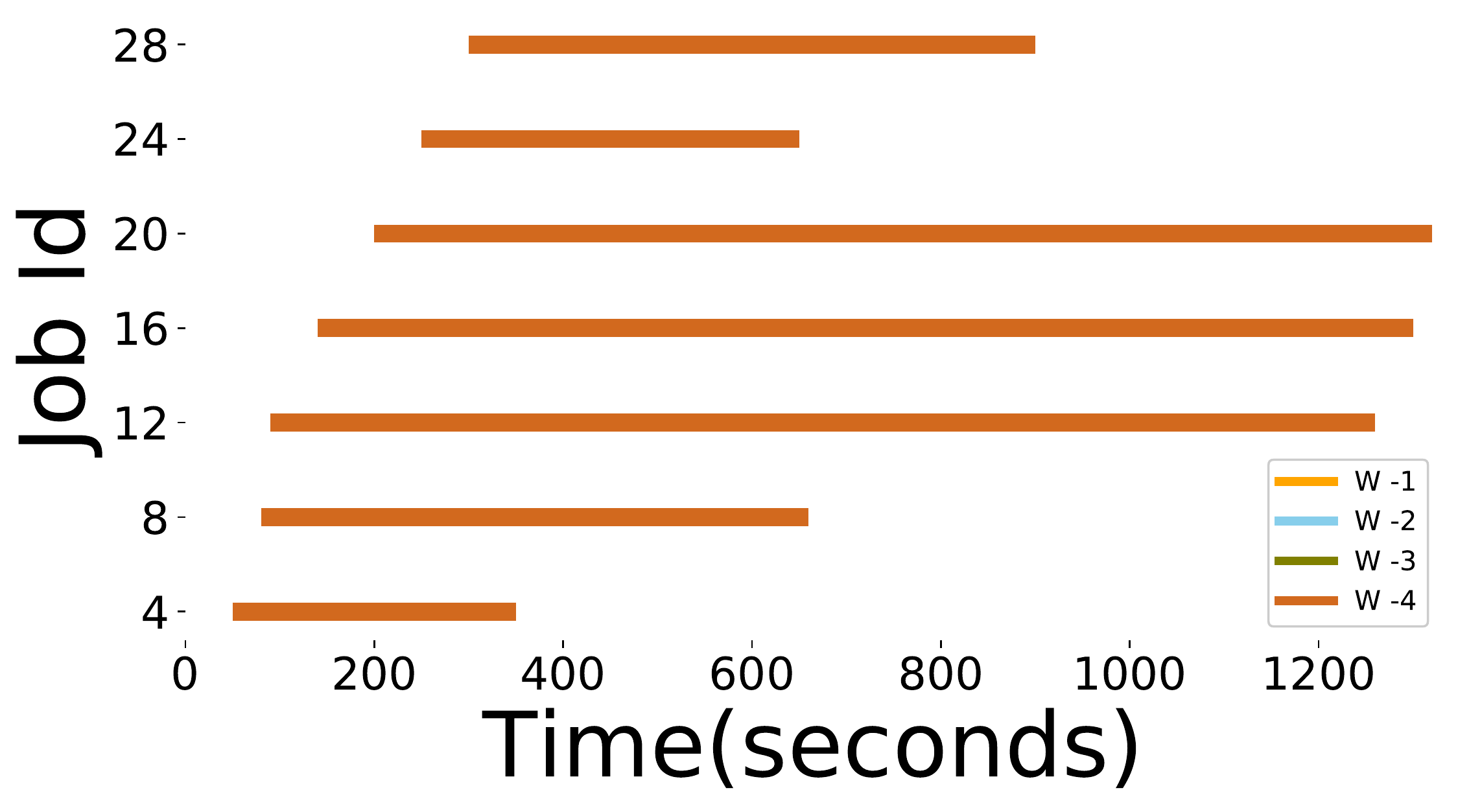}
	\caption{Worker-4}
      \label{fig:random30:ds:node4}
      \end{subfigure} %
\caption{Workload distribution for 30-job experiment with $DS$}  
\label{dist:ds}               
\end{figure*}

\begin{figure*}[ht]
   \centering
         \begin{subfigure}[b]{0.23\textwidth}
\centering
         \includegraphics[width=\textwidth]{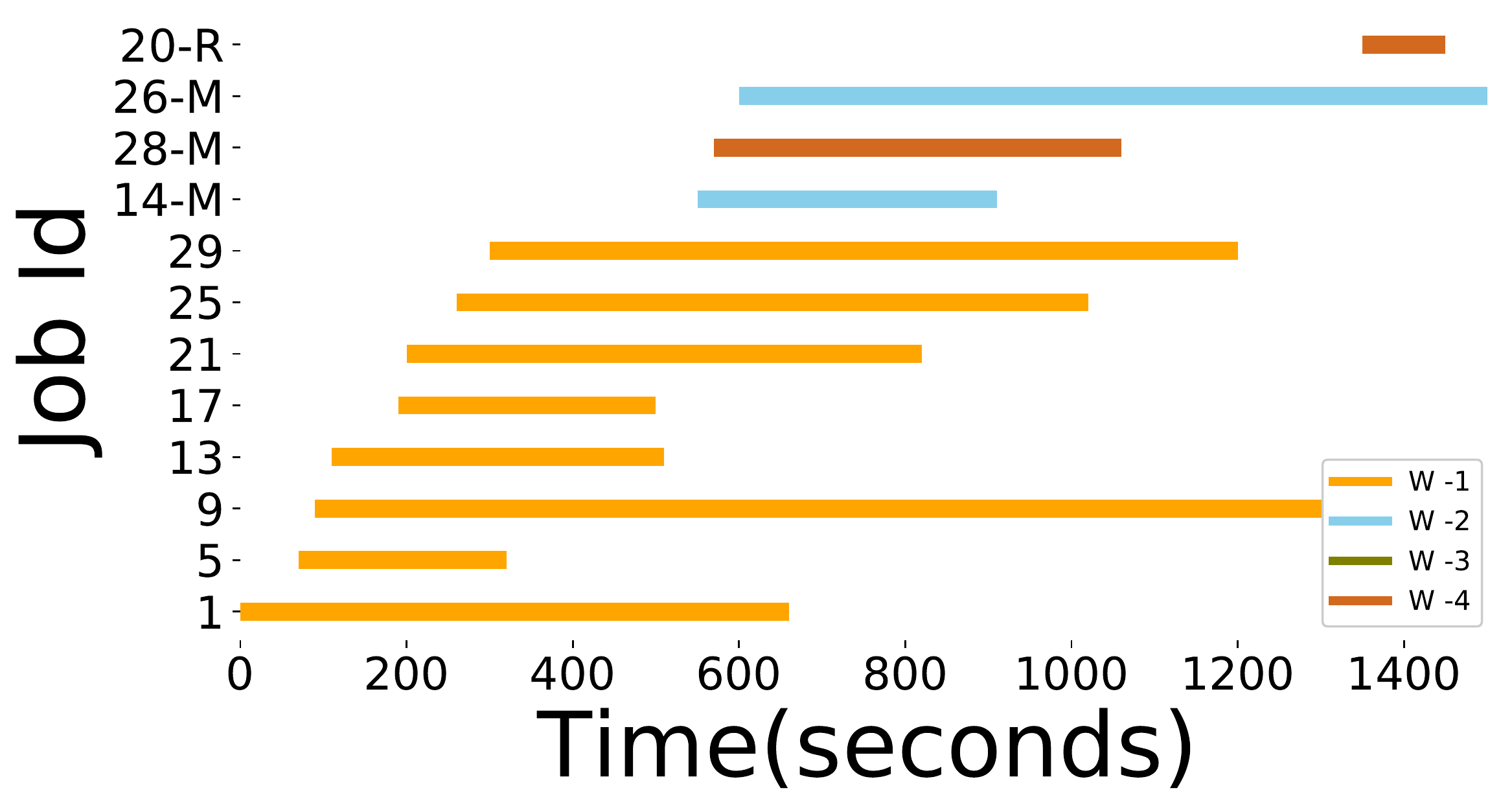}
\caption{Worker-1}
      \label{fig:random30:specon:node1}
      \end{subfigure} 
      \begin{subfigure}[b]{0.23\textwidth}
\centering
         \includegraphics[width=\textwidth]{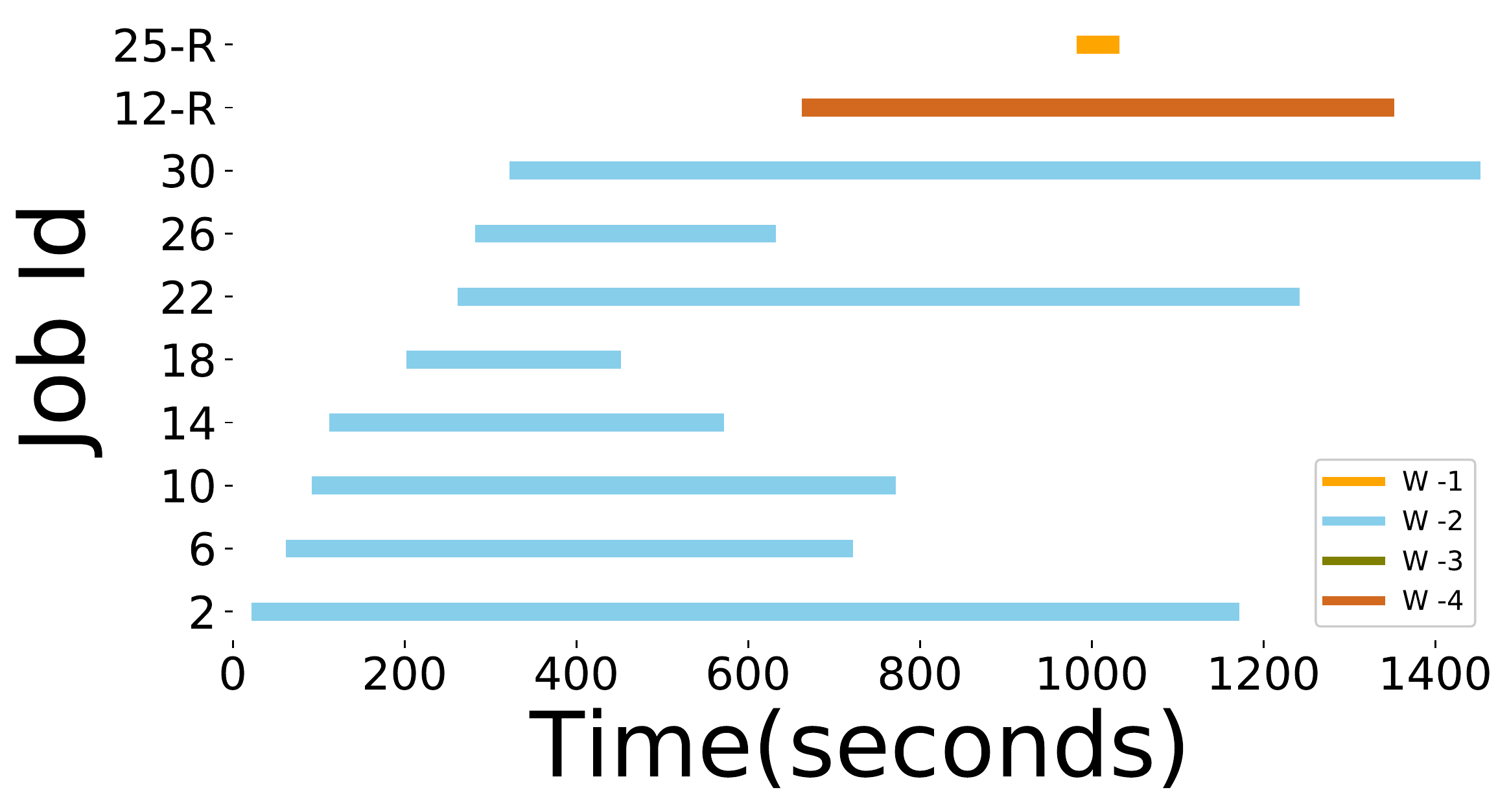}
\caption{Worker-2}
      \label{fig:random30:specon:node2}
      \end{subfigure} %
      \begin{subfigure}[b]{0.23\textwidth}
\centering
         \includegraphics[width=\textwidth]{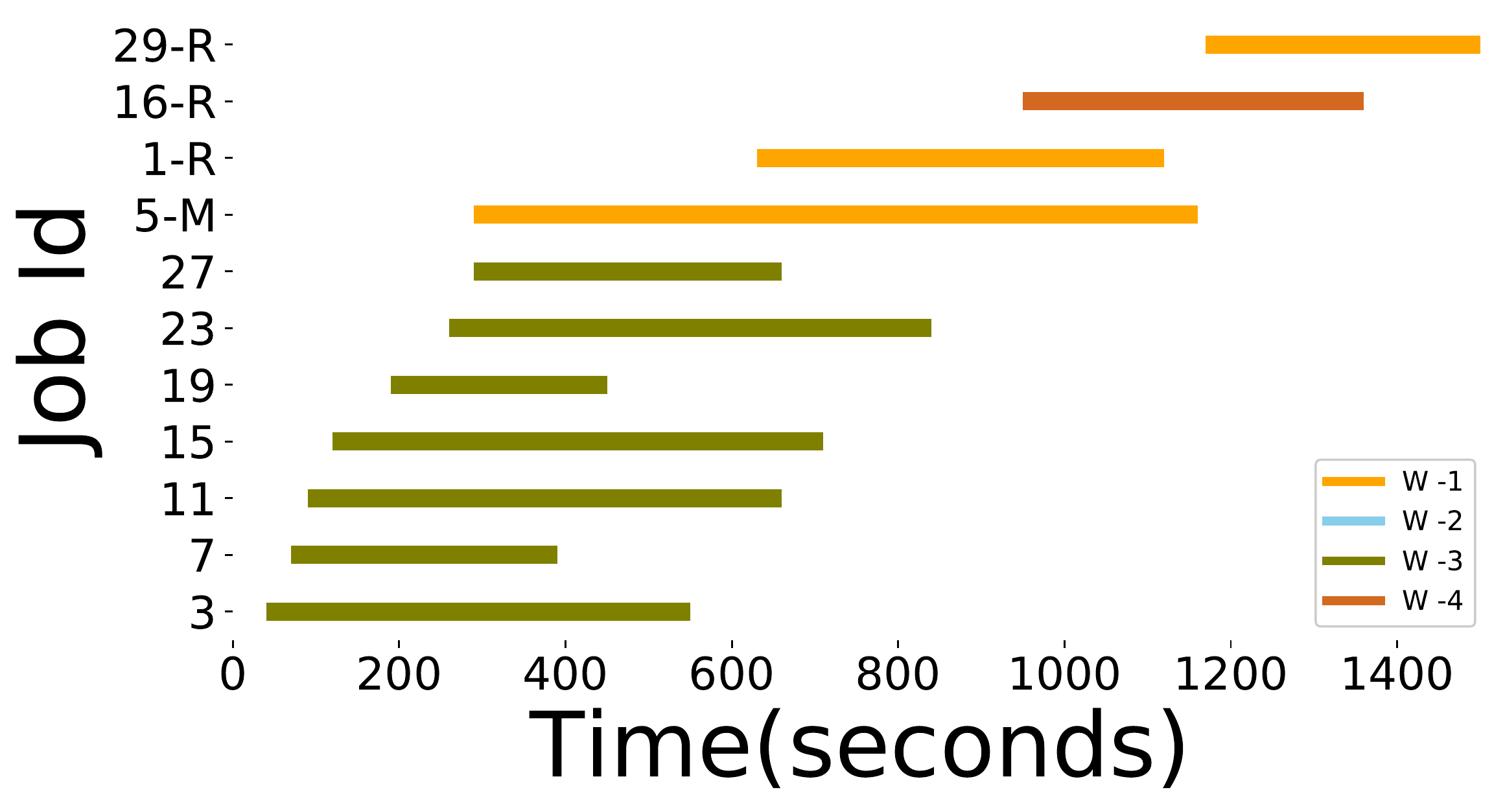}
\caption{Worker-3}
      \label{fig:random30:specon:node3}
      \end{subfigure} %
      \begin{subfigure}[b]{0.23\textwidth}
	\centering
         \includegraphics[width=\textwidth]{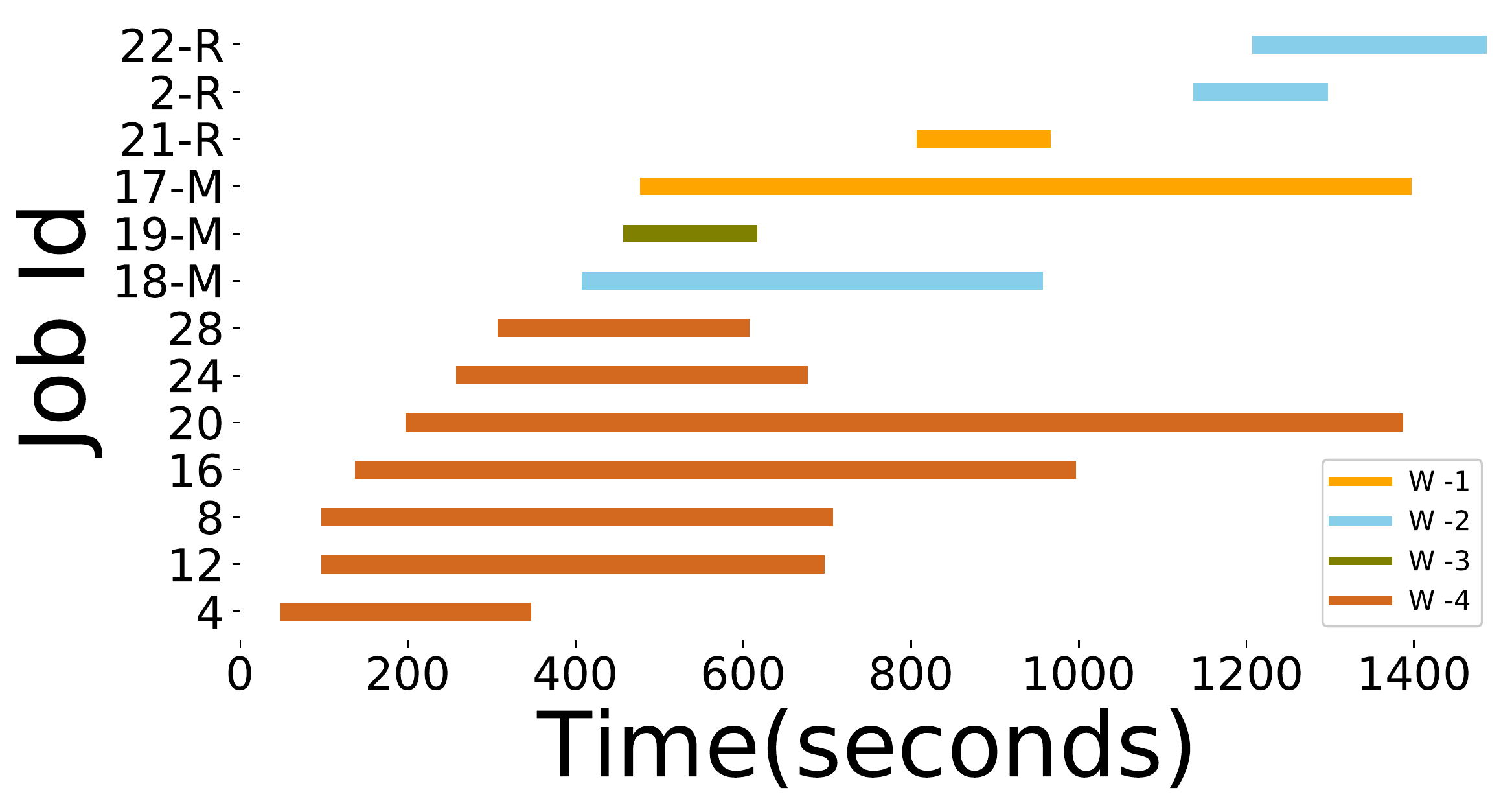}
	\caption{Worker-4}
      \label{fig:random30:specon:node4}
      \end{subfigure} %
\caption{Workload distribution for 30-job experiment with \sol}  
\label{dist:specon}               
\end{figure*}

\begin{figure*}[ht]
   \centering
         \begin{subfigure}[b]{0.23\textwidth}
\centering
         \includegraphics[width=\textwidth]{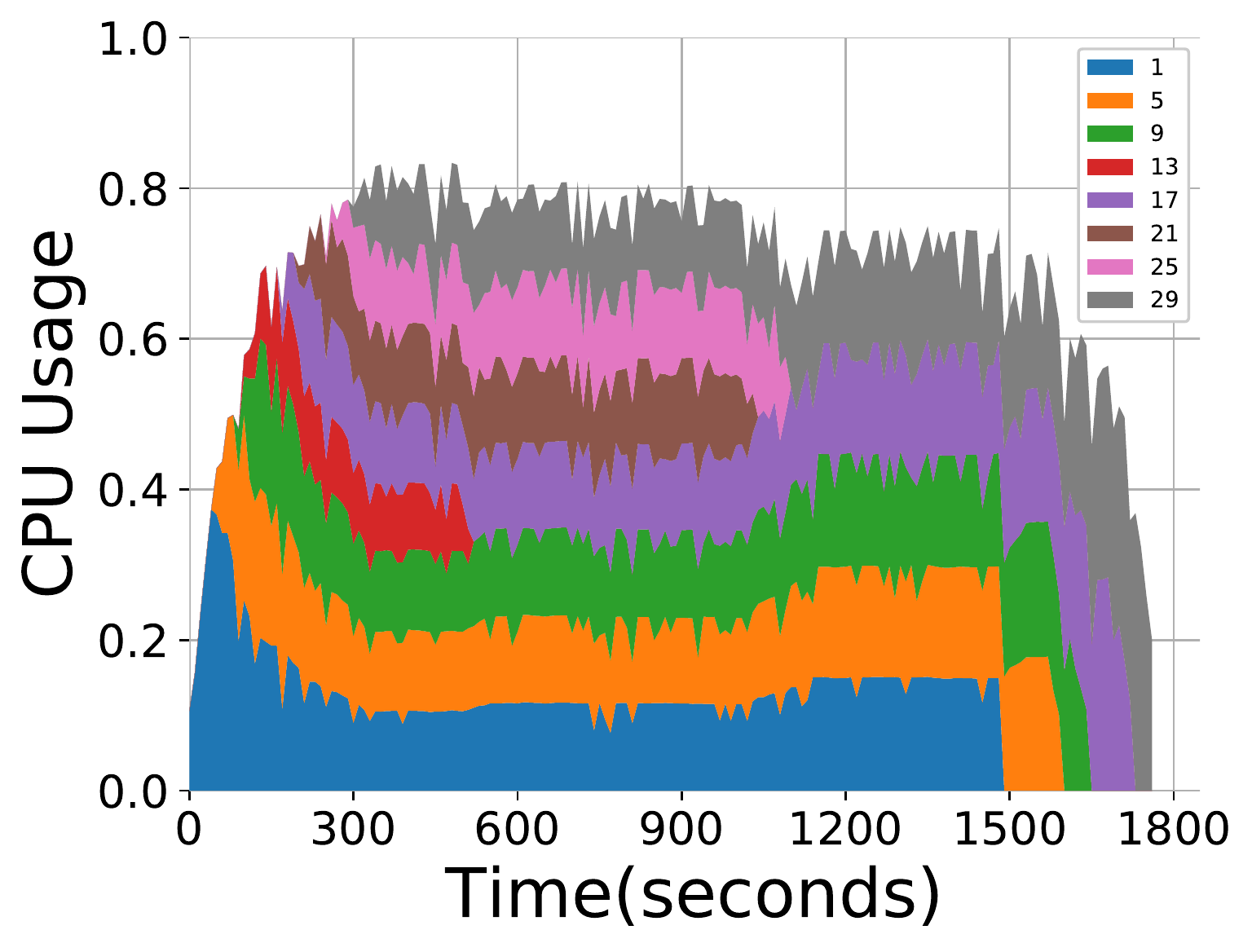}
\caption{Worker-1}
      \label{fig:random30:ds:cpu1}
      \end{subfigure} 
      \begin{subfigure}[b]{0.23\textwidth}
\centering
         \includegraphics[width=\textwidth]{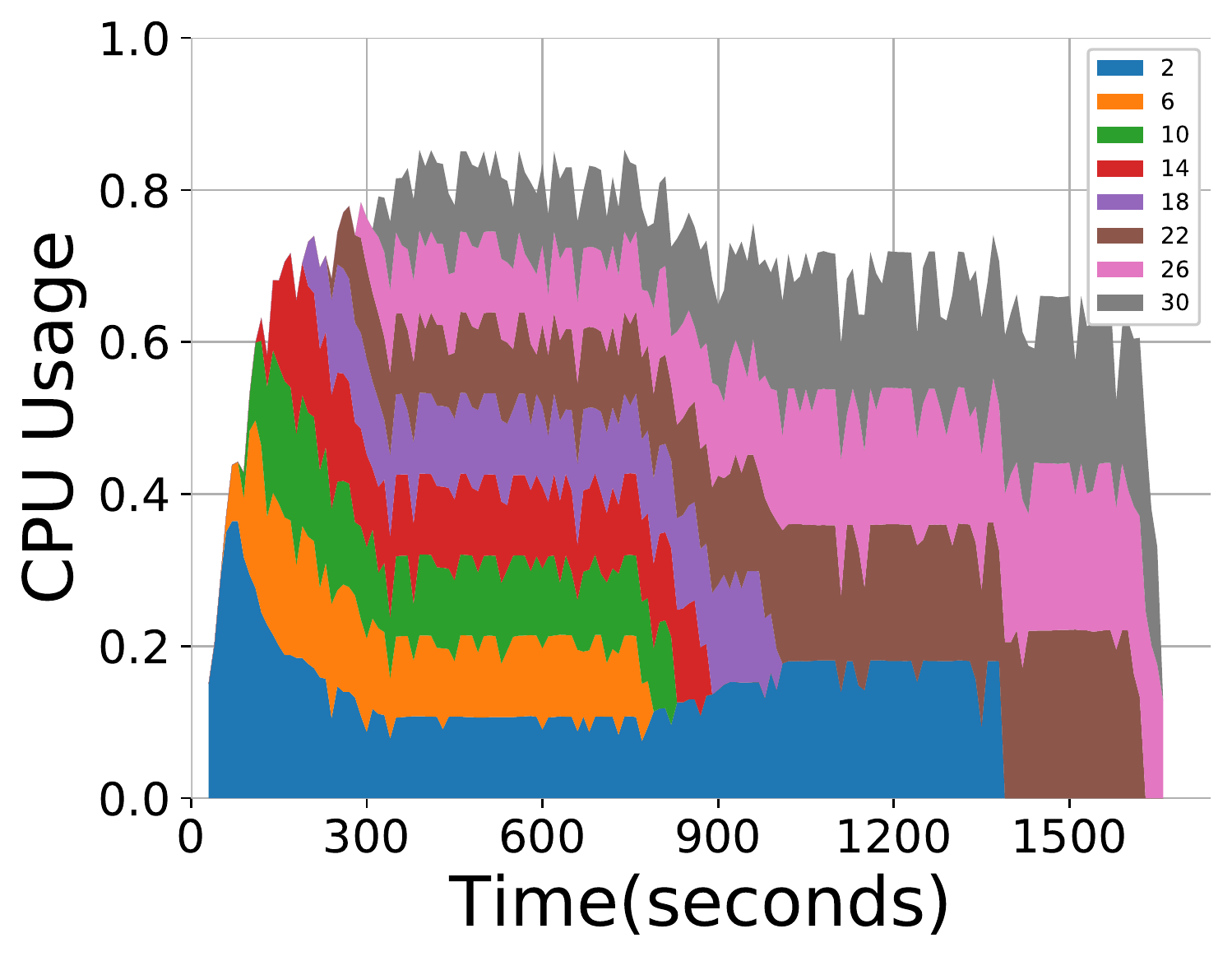}
\caption{Worker-2}
      \label{fig:random30:ds:cpu2}
      \end{subfigure} %
      \begin{subfigure}[b]{0.23\textwidth}
\centering
         \includegraphics[width=\textwidth]{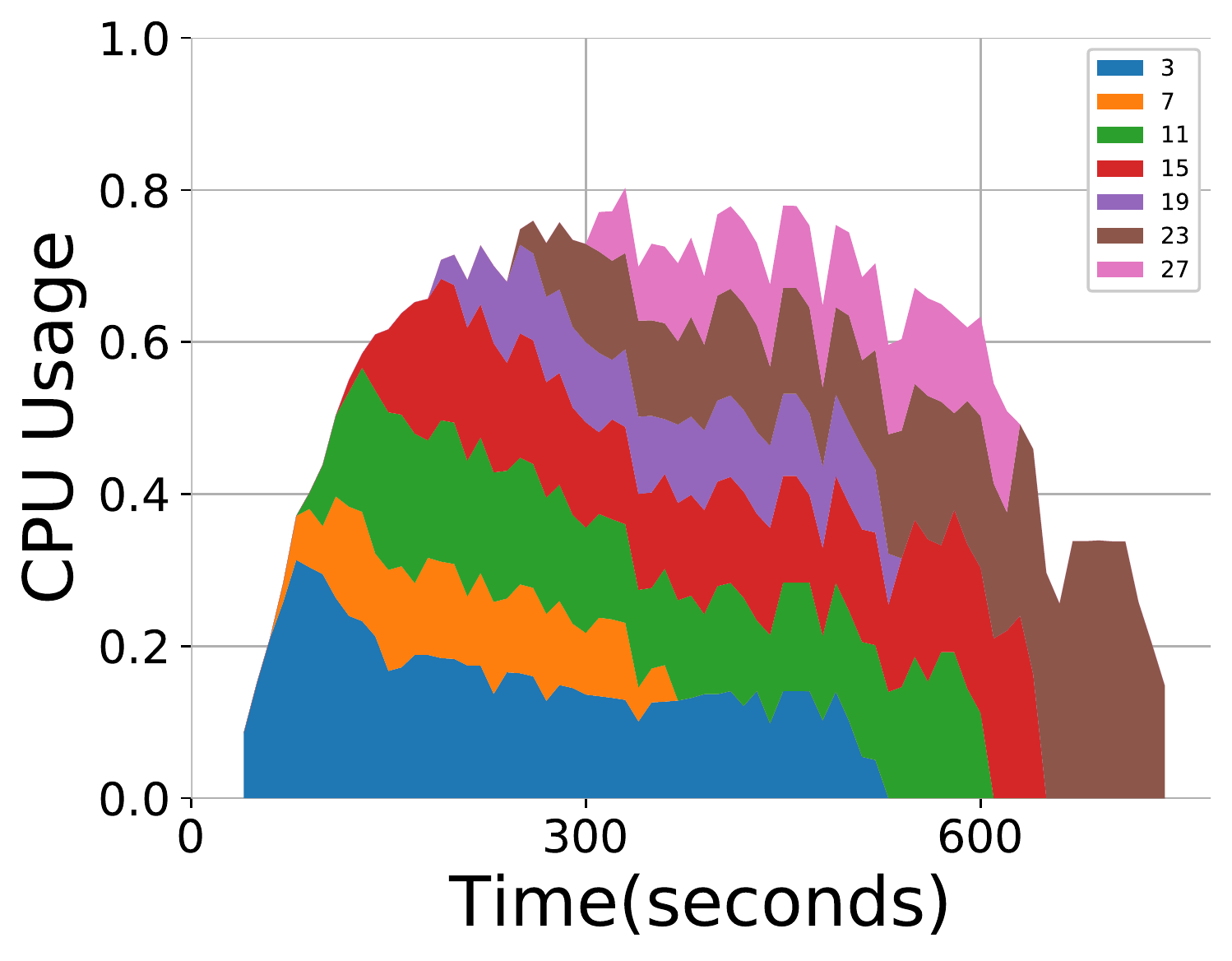}
\caption{Worker-3}
      \label{fig:random30:ds:cpu3}
      \end{subfigure} %
      \begin{subfigure}[b]{0.23\textwidth}
	\centering
         \includegraphics[width=\textwidth]{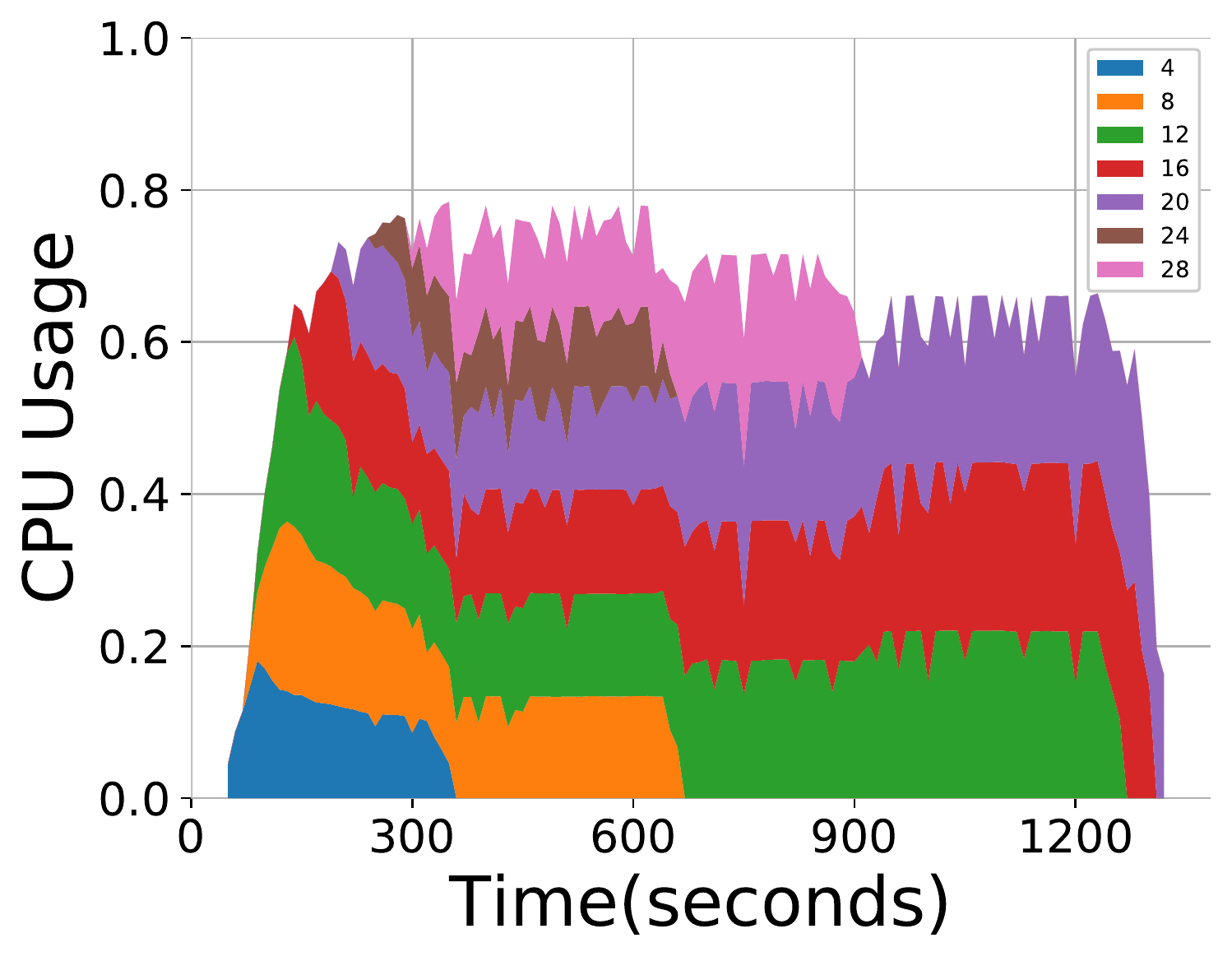}
	\caption{Worker-4}
      \label{fig:random30:ds:cpu4}
      \end{subfigure} %
\caption{CPU usage for 30-job experiment with $DS$}  
\label{cpu:ds}               
\end{figure*}

\begin{figure*}[ht]
   \centering
         \begin{subfigure}[b]{0.23\textwidth}
\centering
         \includegraphics[width=\textwidth]{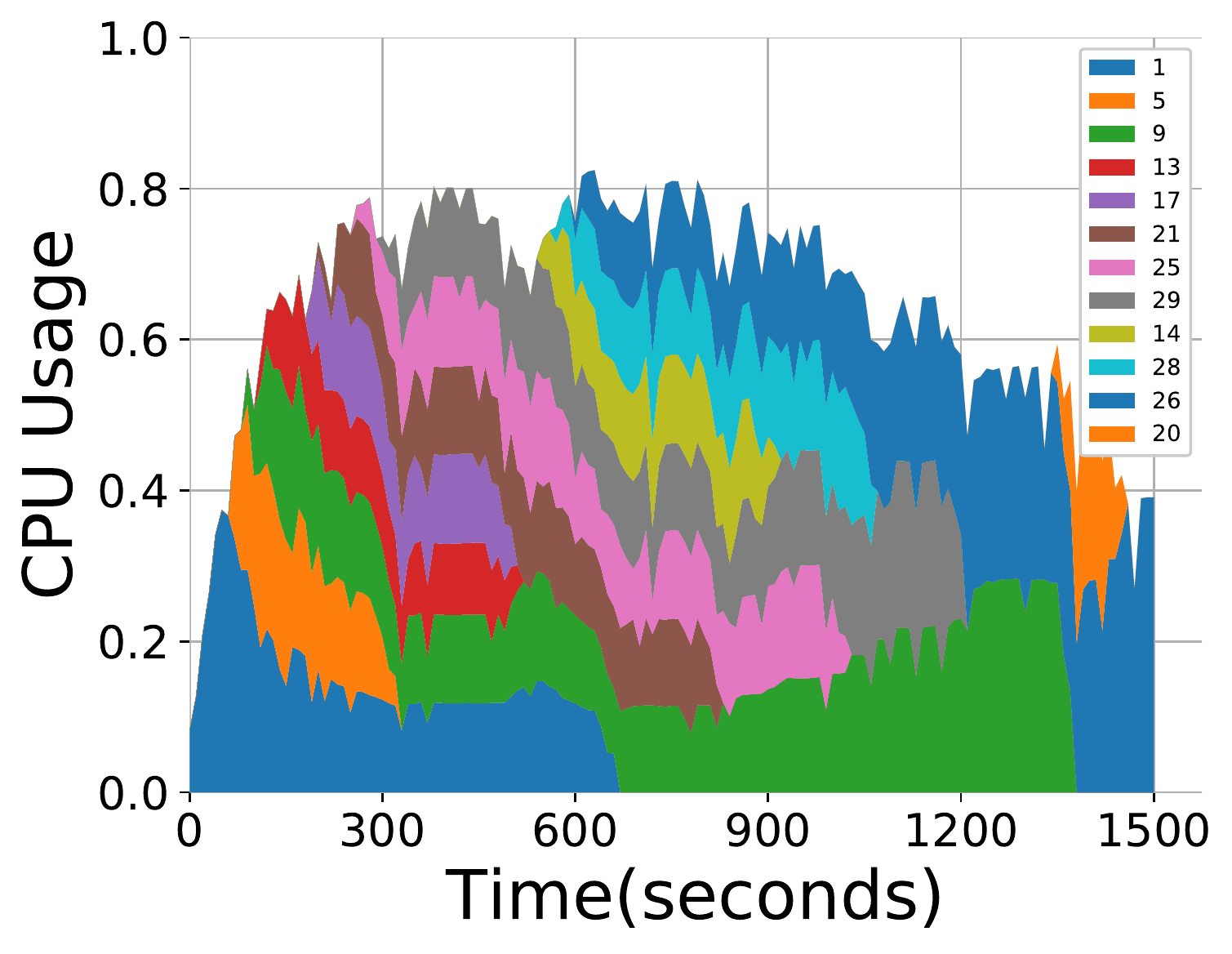}
\caption{Worker-1}
      \label{fig:random30:specon:cpu1}
      \end{subfigure} 
      \begin{subfigure}[b]{0.23\textwidth}
\centering
         \includegraphics[width=\textwidth]{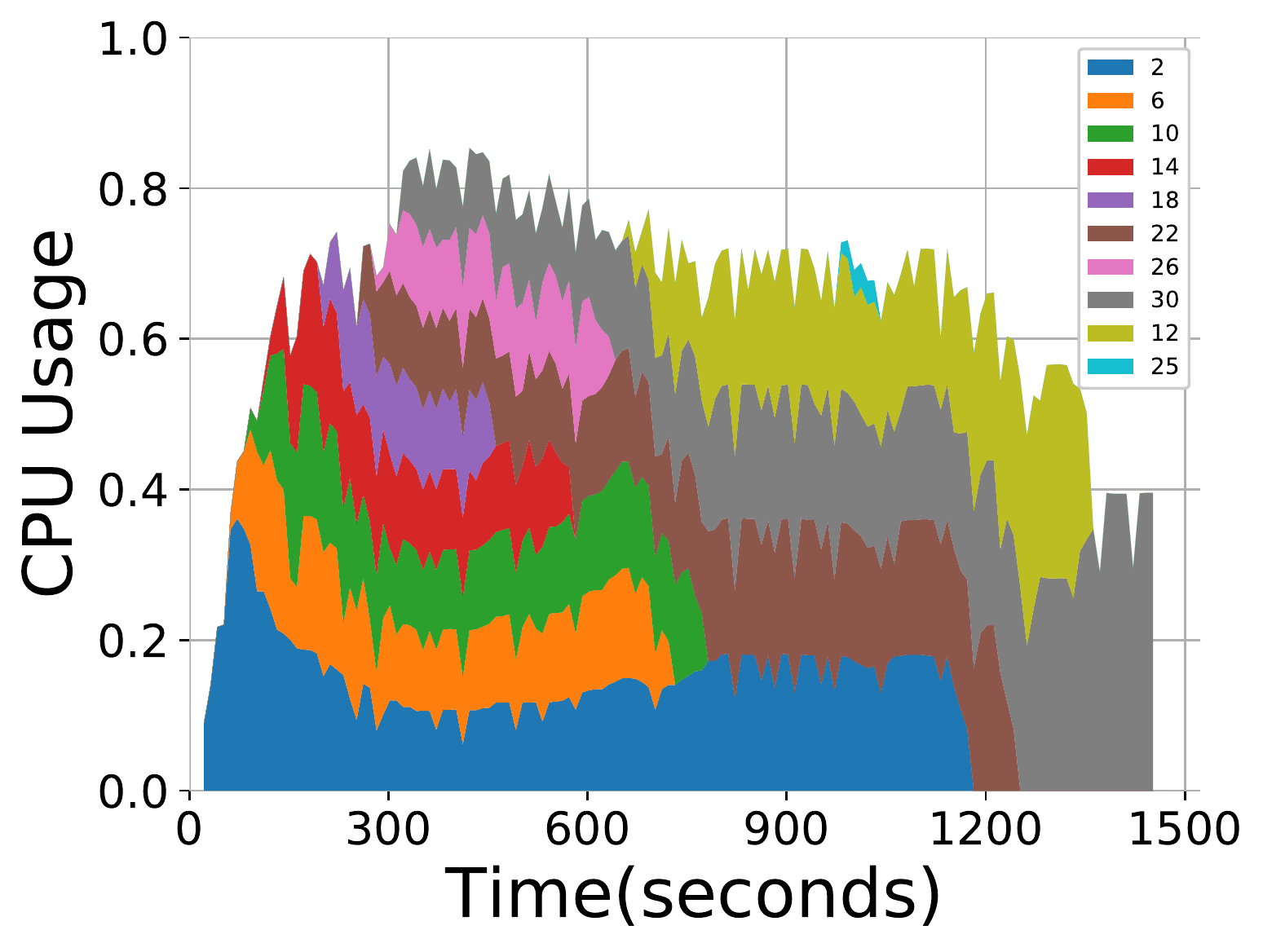}
\caption{Worker-2}
      \label{fig:random30:specon:cpu2}
      \end{subfigure} %
      \begin{subfigure}[b]{0.23\textwidth}
\centering
         \includegraphics[width=\textwidth]{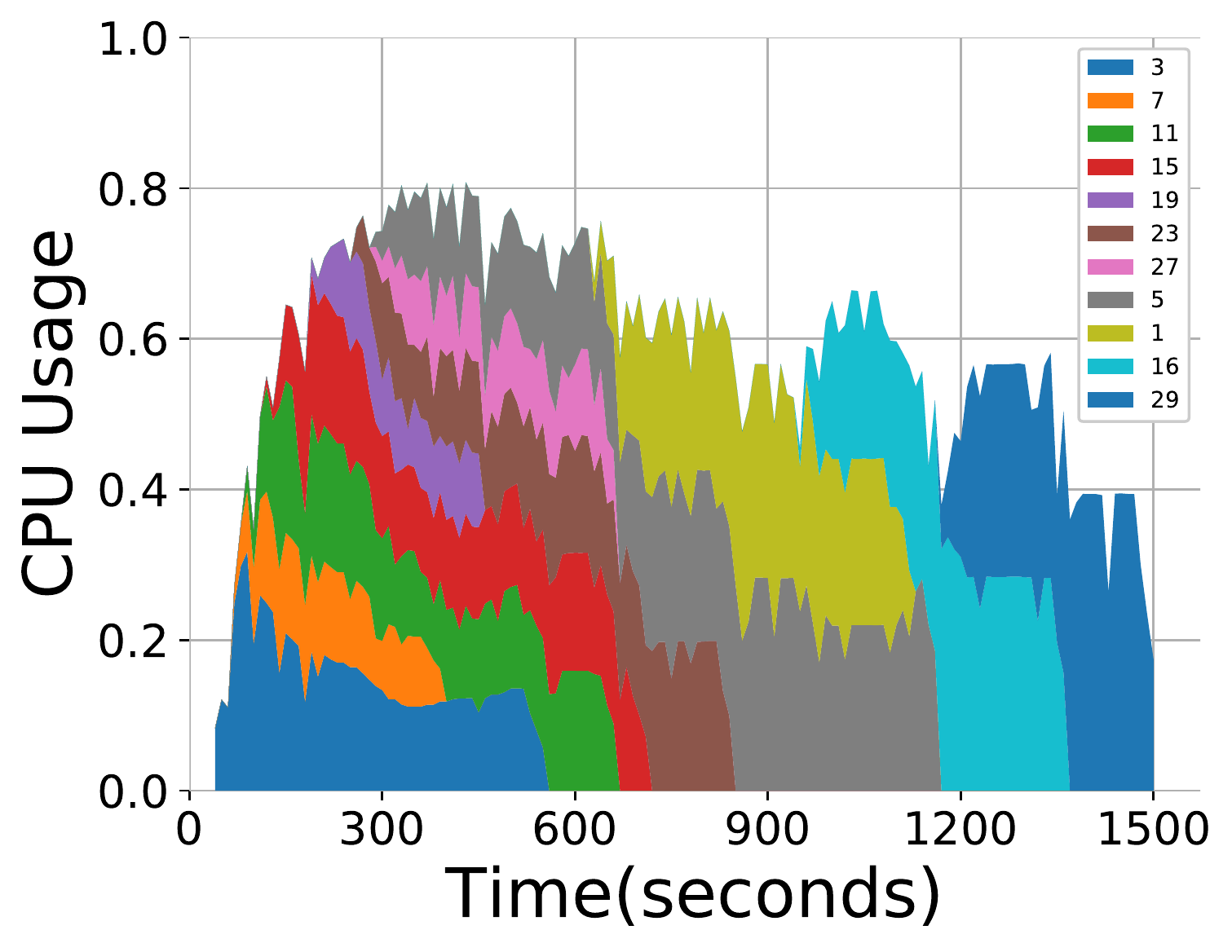}
\caption{Worker-3}
      \label{fig:random30:specon:cpu3}
      \end{subfigure} %
      \begin{subfigure}[b]{0.23\textwidth}
	\centering
         \includegraphics[width=\textwidth]{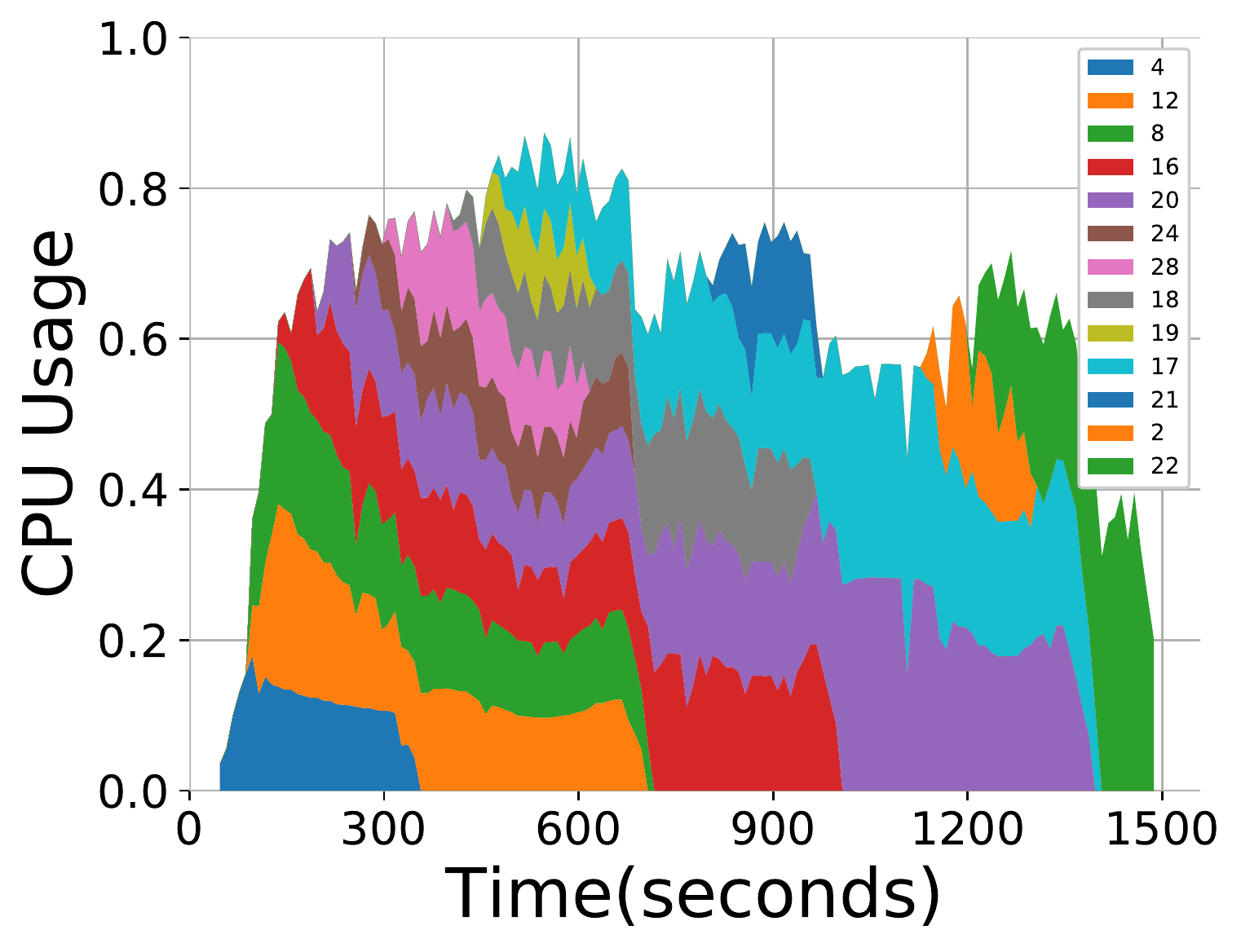}
	\caption{Worker-4}
      \label{fig:random30:specon:cpu4}
      \end{subfigure} %
\caption{CPU usage for 30-job experiment with \sol}  
\label{cpu:specon}               
\end{figure*}

Next, we dig into the details of the experiment.
Fig.~\ref{dist:ds} and Fig.~\ref{dist:specon} illustrate the container distribution over time on each of the workers.
Comparing Fig.~\ref{dist:ds} with Fig.~\ref{dist:specon}, 
we have two observations. (1) As described in the experiment setup,
both $DS$ and \sol~ use a fixed initial container placement for fairness. Therefore, 
container 1, 2, 3, 4, as shown on the figures, is hosted by Worker-1,-2,-3,-4 respectively. 
Then, container 5, 6, 7, 8, rotate to another around of workers.
(2) We can clearly see that none of the containers get migrated. With the default scheduler ($DS$),
the migration only happens when the hosting worker is out of resource or other unhealthy condition. $DS$ fails to 
react based on the job's progress. 
Although, the fixed initial container placement ensures a balance at the beginning, the default scheduler fails to 
take characteristics of loss functions and resource utilization into account. After the initial placement, it lacks a 
mechanism to discover and readjust the system.

With \sol, however, the containers can be migrated according to their training progresses and system-wide workload distribution.
In Fig.~\ref{dist:specon}, we mark the job that is migrated due to its slow-growing progress as "Job-ID-M" and 
"Job-ID-R" indicates that this job is migrated as an reaction of system-wide rebalancing.
For example, Job-14 was was originally running on Worker-2 and it was added to $CC$ category at 537.3s by Algorithm~\ref{alg:1}.
Then, it is delivered to Algorithm~\ref{alg:2} to decide its new host. At this particular moment, the scores of
Worker-1,-2,-3,-4 were 63, 88, 92, 72. Therefore, Job-14 was migrated to Worker-1 and the released resource can be utilized by other jobs on this worker. This migration, together with Job-26's migration (at time 592.1s), benefit Job-2, Job-6, Job-10, Job-30, which all achieve reduction on the completion time.
In addition, for Job-5 on Worker-1, it was added to $CC$ at 286.8s and reassigned to Worker-3, which has 7 active jobs, but 4 of them are in $CC$ already. 
On Worker-3, several of them finished (Job-3, -7, -11, -23, -27) and released resources so that Job-5 can utilize that results in 
a faster completion time.

Furthermore, when the system is unbalanced, Algorithm 3 is executed to re-distribute the workload. For example, Job-1 was originally running on Worker-1 and, at time 620s, the system was unbalanced due to finished job, such as Job-3, Job-7, Job-21.
\sol~ captured this moment and redistributed Job-1 on Worker-1 to Worker-3. This migration ease the resource competition
on Worker-1, which has 8 active containers at the time. It benefits both Job-1, which has been moved to a worker with less workload, and others that can utilize the released resources.



Fig.~\ref{cpu:ds} and Fig.~\ref{cpu:specon} plot the comparison of CPU usage between $DS$ and \sol. 
The resource usage tops at around 80\% since Kubernetes reserves part of them for the system itself.
The resource utilization rates are similar between them. Due to migration and re-distribution, \sol~
presents more jobs on the figures. 



\subsection{Larger cluster}


\begin{figure*}[ht]
\vspace{-0.15in}
\centering
         \includegraphics[width=0.9\linewidth]{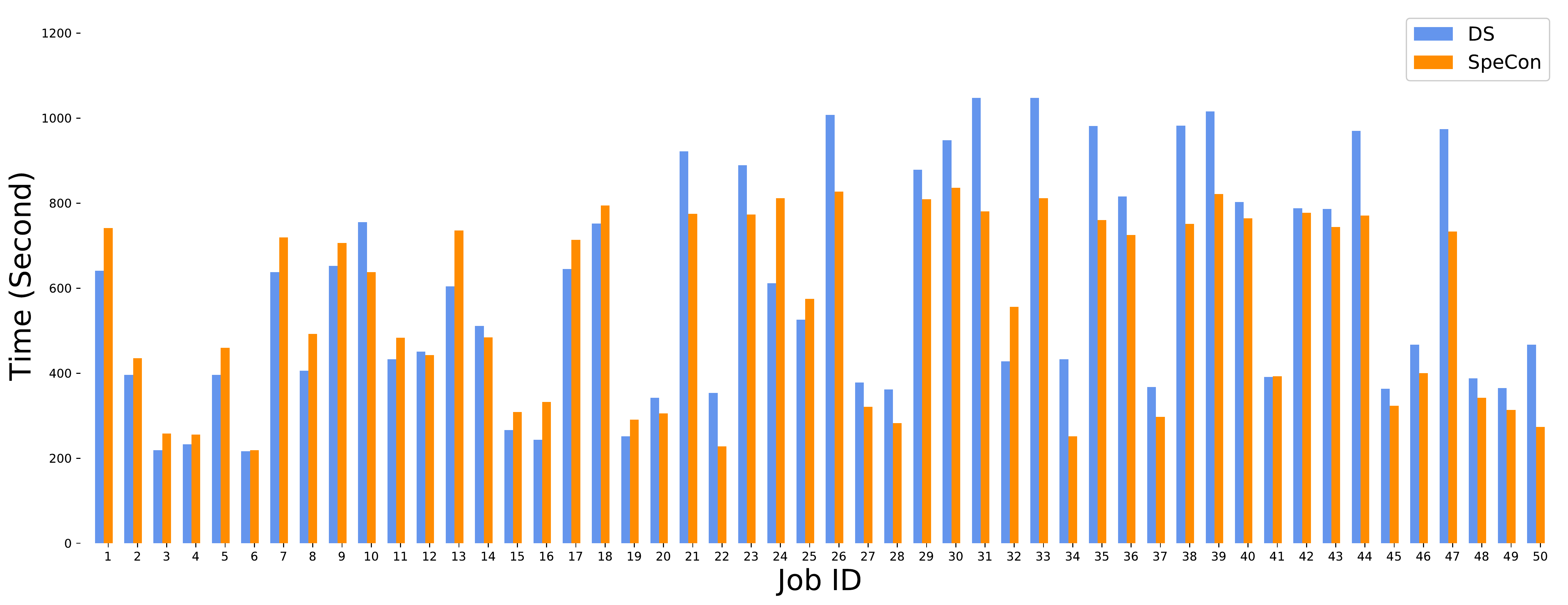}
\caption{Random schedule [0, 1200s] with 50 random jobs}
      \label{fig:scale:50}
\end{figure*}

Finally, we conduct experiment with a larger cluster, 1 manager and 8 workers. The more workers 
give more options for \sol, which leads to challenges when reallocating containers.
Fig.~\ref{fig:scale:50} present the experiment with 50 random jobs running in the larger cluster.
Overall, \sol~ records 30 out of 50 jobs (60.0\%) are completing faster than $DS$. 
The largest gain is found on Job-50, which reduces from 467.5s to 273.4s that is a 41.5\% reduction.
The average completion time reduces 7.2\% from 596.3s to 553.0s. Additionally, the makespan value 
reduces from 2070s to 1840s, 11.1\%.



\section{Conclusion}

We propose \sol~ in this work. It aims to facilitate
speculative container scheduling for deep learning training jobs at runtime. 
We presented the detailed system design
of \sol, and conducted extensive experiments with 5 different
deep learning models on 2 frameworks, Pytorch and Tensorflow,
in a cloud computing environment. 
Our experimental results have
proven the effectiveness of \sol. Specifically, compared to a
default scheduler, it has achieved significant performance boost in the presence of various deep learning workloads, by up
to 41.5\% reduction in completion time for an individual job and 14.8\% in average.
Furthermore, \sol~ achieves improvement on makespan for up to 24.7\%.

\bibliographystyle{IEEEtran}
\bibliography{routing}

\end{document}